\documentclass{aa}  

\usepackage{graphicx}
\usepackage{txfonts}
\usepackage{subcaption}
\newcommand{\teff}{$T_{\rm eff}$}

\newcommand{\msun}{${\rm M}_\odot$}
\newcommand{\lsun}{${\rm L}_\odot$}
\newcommand{\rsun}{${\rm R}_\odot$}

\newcommand{\msunyr}{${\rm M}_\odot {\rm yr}^{-1}$}

\def\kms{km~s$^{-1}$}

\def\teff{$T_{\rm eff}$}
\def\logg{$\log g$}

\def\vsini{$\rm{v}\sin i$}
\def\prot{${\rm P}_{rot}$}
\def\vrad{${\rm V}_{r}$}
\def\vmic{${\rm v}_{mic}$}

\def\ha{${\rm H}\alpha$}
\def\hb{${\rm H}\beta$}

\def\ewha{$EW({\rm H}\alpha$)}

\def\macc{$\dot{M}_{\rm acc}$}
\def\lacc{$L_{\rm acc}$}

\def\rstar{${\rm R}_\star$}
\def\rrad{${\rm R}_{rad}$}
\def\mstar{${\rm M}_\star$}
\def\mrad{${\rm M}_{rad}$}
\def\lstar{${\rm L}_\star$}
\def\prot{${\rm P}_{rot}$}

\def\rcor{$r_{cor}$}
\def\rsub{$r_{sub}$}
\def\rmag{$r_{mag}$}

\begin{document}

   \title{Magnetospheric accretion in the intermediate-mass T Tauri star HQ~Tau\thanks{Based on observations obtained at the Canada-France-Hawaii Telescope (CFHT) which is operated by the National Research Council of Canada, the Institut National des Sciences de l'Univers of the Centre National de la Recherche Scientifique of France, and the University of Hawaii. }}

   \author{K.~Pouilly \inst{1}
          \and
          J. Bouvier \inst{1}
          \and
          E.~Alecian\inst{1}
          \and
          S.H.P.~Alencar\inst{2}
          \and
          A.-M.~Cody\inst{3}
          \and
          J.-F.~Donati\inst{4}
          \and
          K.~Grankin\inst{5}
           \and
          G.A.J.~Hussain\inst{6}
          \and
          L.~Rebull\inst{7}
          \and
          C.P.~Folsom\inst{4}
          }
    
   \institute{Univ. Grenoble Alpes, CNRS, IPAG, 38000 Grenoble, France\\
              \email{kim.pouilly@univ-grenoble-alpes.fr}
              \and
              Departamento  de  Fisica - ICEx - UFMG,  Av.  Ant\^onio  Carlos  6627,  30270-901  Belo  Horizonte,  MG,  Brazil
              \and
              Bay Area Environmental Research Institute, 625 2nd St Ste. 209, Petaluma, CA 94952
              \and
              Univ. de Toulouse, CNRS, IRAP, 14 avenue Belin, 31400 Toulouse, France
              \and
              Crimean Astrophysical Observatory, Nauchny, Crimea 298409
              \and
              European Southern Observatory, Karl-Schwarzschild-Str. 2, 85748 Garching bei M\"unchen, Germany
              \and
              Infrared Science Archive (IRSA), IPAC, California Institute of Technology, 1200 E. California Blvd, MS 100-22, Pasadena, CA 91125 USA
                }

   \date{Received 3 April 2020; Accepted 3 August 2020}

 
  \abstract
   {Classical T Tauri stars (cTTs) are pre-main sequence stars surrounded by an accretion disk. They host a strong magnetic field, and both
   magnetospheric accretion and ejection processes develop as the young magnetic star interacts with its disk. 
   Studying this interaction is a major goal toward understanding the properties of young stars and their evolution.}
   {The goal of this study is to investigate the accretion process in the young stellar system HQ Tau, an intermediate-mass T Tauri star (1.9~\msun).  }
   {The time variability of the system is investigated both photometrically, using Kepler-K2 and complementary light curves, and from a high-resolution spectropolarimetric time series obtained with ESPaDOnS at CFHT. }
   {The quasi-sinusoidal Kepler-K2 light curve exhibits a period of 2.424~d, which we ascribe to the rotational period of the star. The radial velocity of the system shows the same periodicity, as expected from the modulation of the photospheric line profiles by surface spots. A similar period is found in the red wing of several emission lines (e.g., HI, CaII, NaI), due to the appearance of inverse P Cygni components, indicative of accretion funnel flows. Signatures of outflows are also seen in the line profiles, some being periodic, others transient. The polarimetric analysis indicates a complex, moderately strong magnetic field which is possibly sufficient to truncate the inner disk close to the corotation radius, \rcor$\sim$3.5~\rstar. Additionally, we report HQ Tau to be a spectroscopic binary candidate whose orbit remains to be determined. }
   {The results of this study expand upon those previously reported for low-mass T Tauri stars, as they indicate that the magnetospheric accretion process may still operate in intermediate-mass pre-main sequence stars, such as HQ Tau. }


   \maketitle

%

\section{Introduction}

   Classical T Tauri stars (cTTs) are young stellar objects still surrounded by an accretion disk.
   They possess a strong magnetic field that truncates the inner disk at a distance of a few stellar radii above the stellar surface and drives accretion through funnel flows, a process referred to as magnetospheric accretion \citep[see reviews in, e.g.,][]{Bouvier07b, Hartmann16}.
   The kinetic energy of the infalling material is dissipated in a shock at the stellar surface, creating a localized hot spot.
   The star-disk interaction takes place within 0.1 au or less, a scale hardly resolved by current interferometers. An alternative approach to study this compact region is to monitor the variability of the system through contemporaneous photometric, spectroscopic, and spectropolarimetric observing campaigns focused on specific targets. Over the past decade, our group has reported several such studies   \citep[e.g.,][]{Bouvier07a, Donati07, Alencar12, Alencar18, Donati19}. 

   This new study focuses on \object{HQ Tau} (RA=04h35m, Dec.=+22$^{\circ}$50$\arcmin$), a 1.9~\msun\ T Tauri star located in the Taurus star forming region at a distance of 159~pc \citep{Gaia18}. 
   The system is moderately bright (V$\simeq$12.5) \citep{Norton07}, has a K0-K2 spectral type \citep{Nguyen12, Herczeg14}, and is so far considered as single \citep[][but see Section 4 below]{Pascucci15}.
From the SuperWASP campaign,  \cite{Norton07} measured a light curve modulation with a peak-to-peak amplitude of about 2~mag in the V-band and a period of 2.4546 days. \cite{Rodriguez17} analyzed a 9~yr long KELT light curve of the system and reported several long-duration, non-periodic dimming events with an amplitude of about 1.5 mag and lasting for weeks to months, which led them to classify this object as UXor-like \citep{Grinin92}.  \cite{Nguyen12} derived a radial velocity of 16.65 $\pm$ 0.11 \kms, \vsini\ of 48 $\pm$ 2 \kms,  and a 10\% width of the \ha\  line profile of 442 $\pm$ 93 \kms, indicative of ongoing accretion onto the star \citep[see also][]{Duchene17}.
   \cite{Simon16} measured \ewha = 2.22 \AA, from which they derived a mass accretion rate of \macc = 2 $\times10^{-9}$ \msunyr.
More recently, from high-resolution ALMA observations, \cite{Long19} derived an inclination of 53.8 $\pm$ 3.2\degr\ for the circumstellar disk on a scale of $\sim$25 au. HQ Tau is the cTTs with the faintest disk of their Taurus sample in the mm range, and they report a depletion of dust toward the inner regions,  perhaps indicative of an unresolved inner disk cavity. \cite{Akeson19} similarly derived a low disk mass M$_{d}$ = 0.4~10$^{-3}$~\msun\ and mm flux compared to single young stars of similar mass.
   
We resolved to launch a monitoring campaign on this object as a representative of intermediate-mass T Tauri stars (IMTTs), having a mass at the upper range of cTTs.  Moreover, its K2 light curve exhibits a clear and smooth periodicity, which makes this target amenable to the variability approach developed here. Our main goal is to investigate whether the magnetospheric accretion process, which seems to be quite common among low-mass T Tauri stars applies as well to the intermediate-mass range.    In Section \ref{sec:obs}, we describe the photometric and spectropolarimetric datasets used in this study.
In Section \ref{sec:results}, we present the results of the analysis, which includes deriving stellar parameters and investigating the photometric, spectral, and polarimetric variability. We discuss the results in Section \ref{sec:discussion}, where we provide a framework for their interpretation. We also report long-term radial velocity variations, which indicate that the system is a spectroscopic binary.  
We conclude on the presence of the magnetospheric accretion process in this young system of the intermediate-mass group in Section \ref{sec:conclusion}.


\section{Observations}

We briefly describe here the origin of the photometric and spectropolarimetric observations that were obtained during the HQ Tau campaign. We also describe how they were processed. 

\label{sec:obs}
\subsection{Photometry}

    HQ Tau (EPIC 247583818) was observed by Kepler-K2 during Campaign 13, which took place over 80 days from March 8, 2017 to May 23, 2017. 
    The observations were performed in a broad band filter (420-900 nm) with measurements taken at a cadence of 30 minutes. 
    The K2 light curve was reduced by A.-M. Cody \citep{Cody18} and we used the PDC version in this work. 
    
    Additional photometric observations of HQ Tau were secured in the Johnson's VR$_J$I$_J$ filters at the Crimean Astrophysical Observatory (CrAO) from December 22, 2016 to February 15, 2018, on the AZT-11 1.25m telescope equipped with the CCD camera ProLine PL23042. 
    CCD images were bias subtracted and flat-field corrected following a standard procedure. We performed differential photometry between HQ Tau and a non-variable, nearby comparison star, 2MASS J04361769+2247125, whose brightness and colors in the Johnson system are 10.16 (V), 1.61 (V-R)$_J$, and 2.98 (V-I)$_J$.
    A nearby control star of similar brightness, 2MASS J04362344+2252171, was used to verify that the comparison star was not variable. It also  provided an estimate of the photometric rms error in each filter, which amounts to 0.010, 0.012, and 0.011 in the VR$_J$I$_J$ bands, respectively.

\subsection{Spectroscopy and spectropolarimetry}
    High-resolution optical spectropolarimetry was obtained for HQ Tau using the Echelle SpectroPolarimetric Device for the Observation of Stars (ESPaDOnS) \citep{Donati03} at the Canada-France-Hawaii Telescope (CFHT) between October 28, 2017 and November 9, 2017.
    We obtained 14 spectra covering the 370-1000 nm range at a resolution of 68,000, reaching a S/N from 120 to 200 at 731 nm. Each observation consists in the combination of 4 individual spectra taken in different polarization states.
    All the data were reduced using the Libre-ESpRIT package \citep{Donati97}, which provides an intensity Stokes I spectrum and a polarized Stokes V spectrum for each observation. 
    The reduced spectra were normalized using a polynomial fit over selected continuum points \citep[see][]{Folsom16}, which produces a flat continuum, and simultaneously accounts for the spectral order overlap. 
    The journal of observations is provided in Table \ref{tab:log_obs}.
        
    \begin{table}[ht]
    \centering
    	\begin{tabular}{l l l l l}
    	    \hline
    	    Date & HJD & S/N & S/N$_{LSD}$ & $\phi_{rot}$\\\
    	    (2017) & (2,450,000+) &  &  \\
            \hline
    	    28 Oct & 8054.93481 & 165 & 687 & 0.349 \\
    	    29 Oct & 8055.92519 & 180 & 776 & 0.757 \\
    	    30 Oct & 8056.97501 & 156 & 656 & 1.190 \\
    	    31 Oct & 8057.91808 & 121 & 482 & 1.579 \\
    	    01 Nov & 8058.99693 & 170 & 655 & 2.024 \\
    	    02 Nov & 8060.01605 & 180 & 784 & 2.445 \\
    	    03 Nov & 8060.98848 & 191 & 842 & 2.846 \\
    	    04 Nov & 8061.91652 & 204 & 923 & 3.229 \\
    	    05 Nov & 8063.04627 & 169 & 724 & 3.695 \\
    	    07 Nov & 8064.96108 & 192 & 835 & 4.485 \\
    	    08 Nov & 8065.92044 & 140 & 432 & 4.881 \\
    	    08 Nov & 8065.96526 & 153 & 609 & 4.899 \\
    	    08 Nov & 8066.08352 & 180 & 761 & 4.948 \\
    	    09 Nov & 8066.88036 & 174 & 712 & 5.277 \\
  	    \hline
    	\end{tabular}
	\caption{Journal of ESPaDOnS observations. The columns list the date of observation, the Heliocentric Julian Date, the S/N by spectral resolution element at 731 nm, the effective S/N of the Stokes V LSD profiles, and the rotational phase.}
    \label{tab:log_obs}
    \end{table}

    
\section{Results}

We describe in the following subsections the analysis of photometric, spectral, and spectropolarimetric variations observed in the HQ Tau system. We also derive the system's fundamental properties. 

\label{sec:results}
\subsection{Photometric variability}

\begin{figure}
  \centering
\includegraphics[width=.48\textwidth]{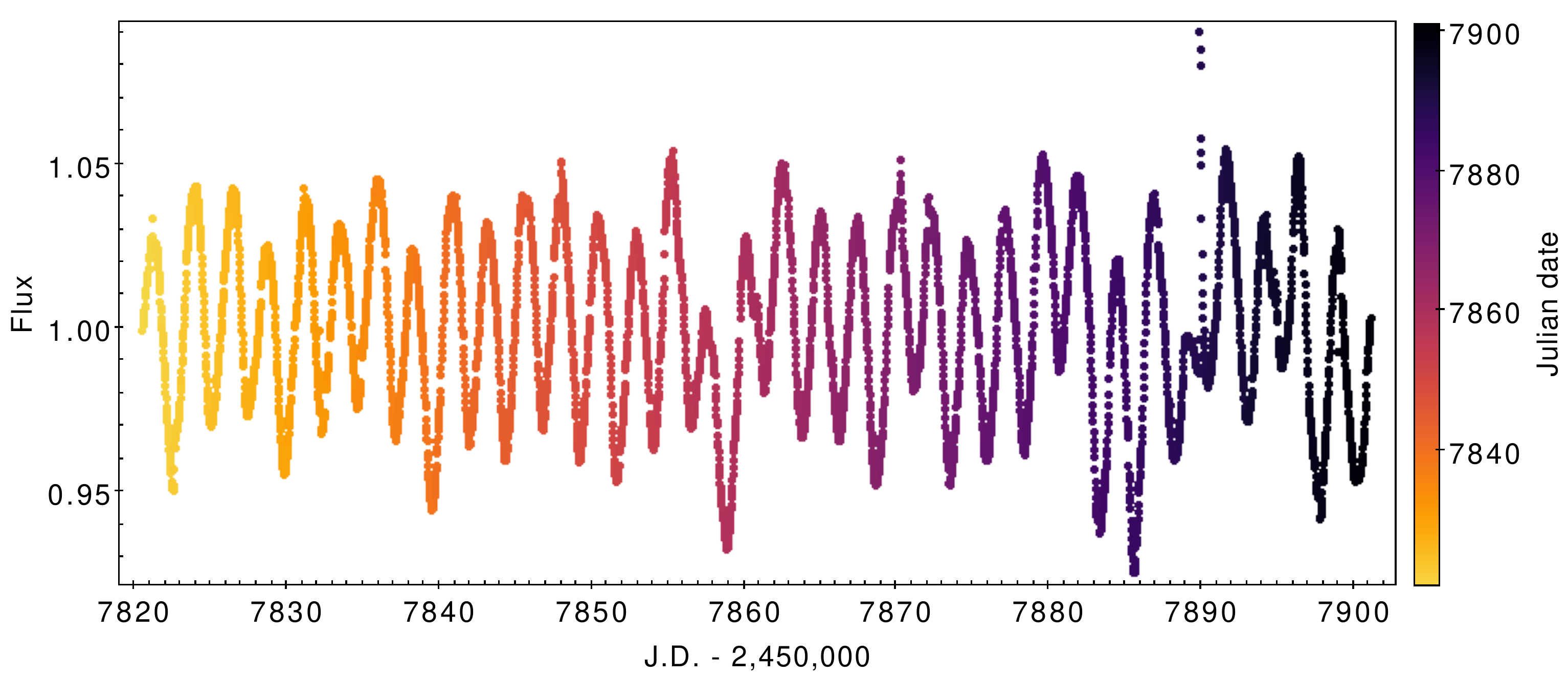}
\includegraphics[width=.48\textwidth]{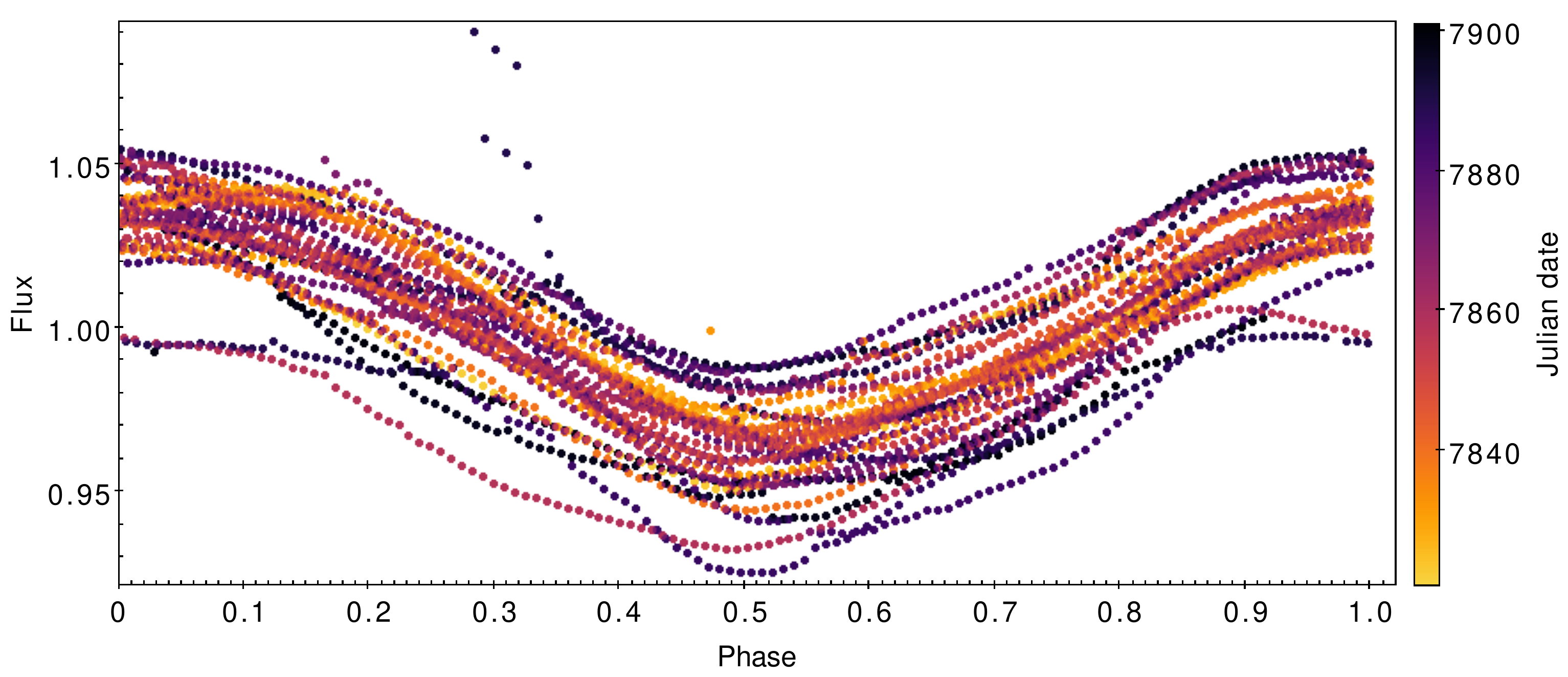}
   \includegraphics[width=.45\textwidth]{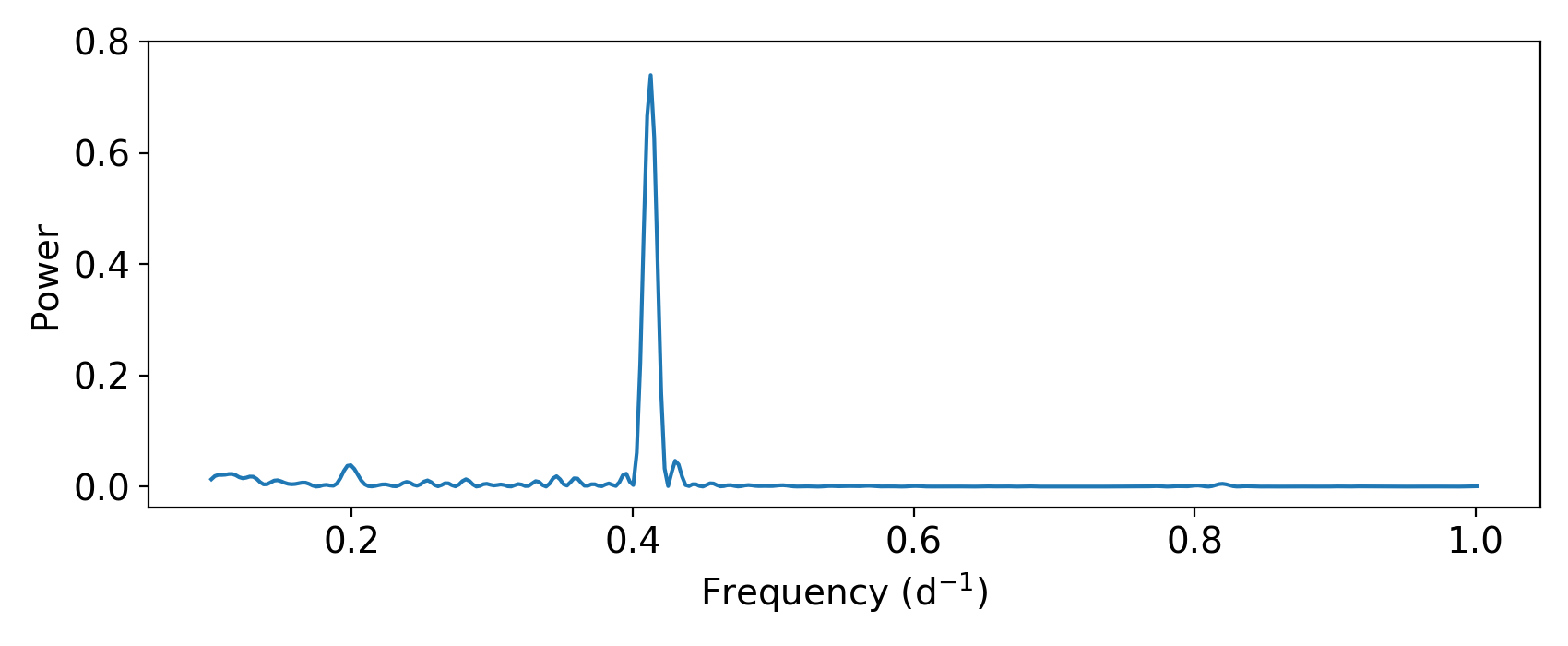}
  \caption{{\it Top:} HQ Tau's detrended Kepler K2 light curve. 
  Low frequency variations have been removed from the original light curve shown in Fig.\ref{fig:lc}. The spike appearing at J.D. 2,457,890.0 is presumably an instrumental defect. 
 {\it Middle:} HQ Tau's light curve folded in phase with P = 2.424~d. The origin of phase is taken as JD 2,457,823.81 to correspond to the epoch of maximum brightness. 
 {\it Bottom:} Lomb-Scargle periodogram obtained from HQ Tau's Kepler K2 light curve. }
   \label{fig:lightcurve}
\end{figure}

   The 80 day-long detrended Kepler-K2 light curve of HQ Tau is shown in Fig. \ref{fig:lightcurve}. It exhibits clear periodic modulation.
   A Lomb-Scargle periodogram analysis \citep{Scargle82,Press89}  yields a period P = 2.424~d, where the period estimate is derived from a Gaussian fit to the periodogram's main peak, with an associated false alarm probability (FAP) of less than 10$^{-4}$, computed following the \cite{Baluev08} approximation. Assuming that the formal error on the period is measured as the $\sigma$ of the Gaussian fit \citep{Vanderplas18}, it would amount to 0.028~d. However, when the K2 light curve is folded in phase, it is clear that the phase coherency between the first and last minima is lost, meaning that their location differs by at least  0.05 in phase, as soon as the period is varied by more than 0.002~d. We therefore use this empirical estimate as the true error on the period.
   The light curve folded in phase with this period (see Fig.\ref{fig:lightcurve}) displays a stable, sinusoidal-like  pattern with a nearly constant amplitude, which we ascribe to the modulation of the stellar flux by a cool surface spot \citep{Herbst94}.
   We therefore assume the photometric period P=2.424 $\pm$ 0.002~d is the rotational period of the star.
   This period is consistent with the P1=2.42320~d period found by \cite{Rebull20} whose work also highlighted a secondary period P2=5.02495~d. We do see a secondary peak in the periodogram of the detrended light curve at a period of 5.04 $\pm$ 0.13~d, with a FAP of less than 10$^{-4}$ (see Fig.~\ref{fig:lightcurve}). We argue below that HQ Tau is probably a binary system, and we cannot exclude that the secondary period is to be associated with the rotation period of the companion.

\begin{figure*}
        \centering
        \includegraphics[width=.95\textwidth]{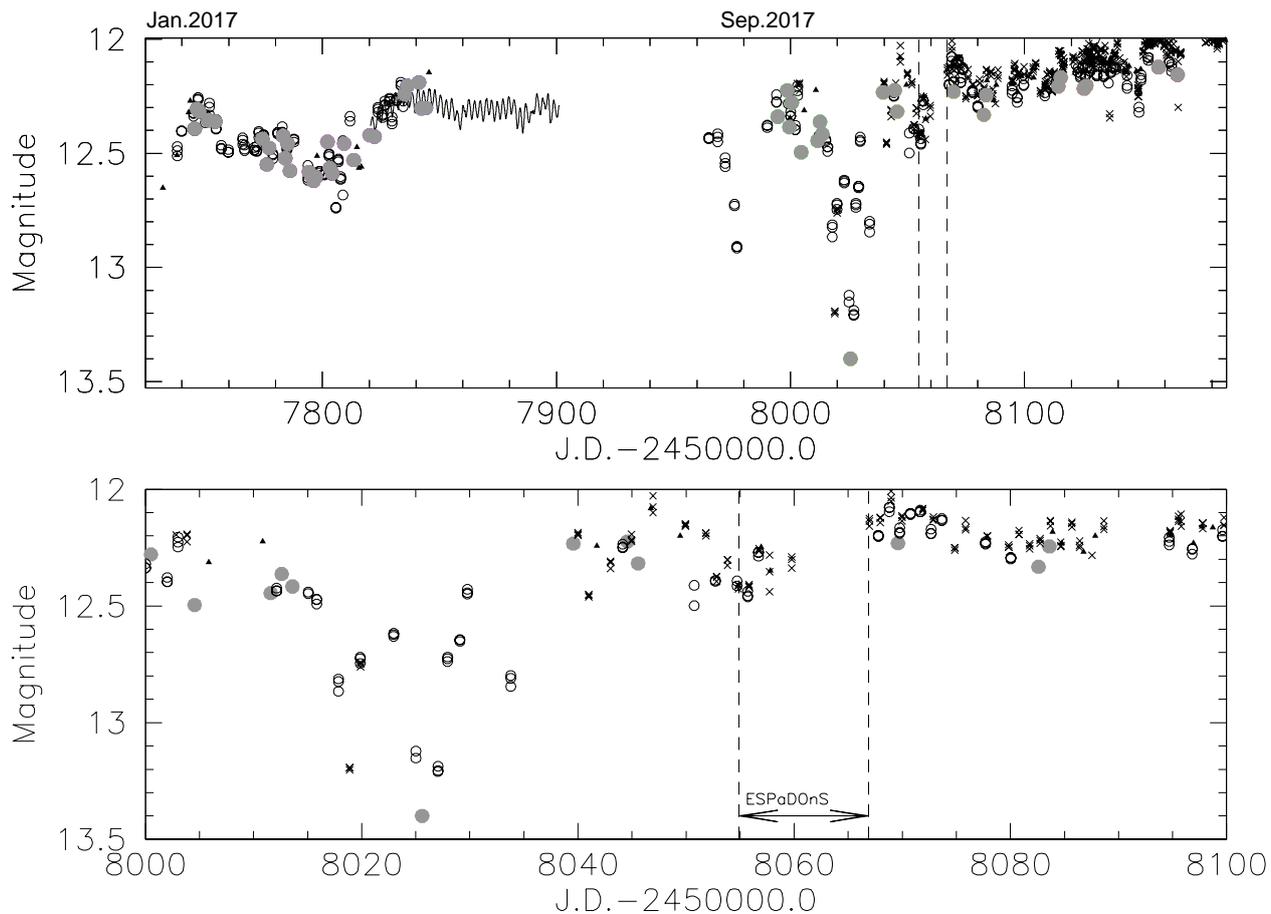}
        \caption{HQ Tau's V-band light curve. CrAO measurements are shown as filled grey circles, the K2 light curve as a continuous line, AAVSO measurements as filled triangles, and the ASAS-SN rescaled g- and V-band datasets as crosses and open circles, respectively. The upper panel shows the light curve over 2 observing seasons while the lower one displays the photometric variability of the source around the CFHT/ESPaDOnS observations. }
        \label{fig:lc}
    \end{figure*}

    Figure \ref{fig:lc} shows the full HQ Tau light curve over 2 epochs, from December 2016 to March 2018, including the K2, before detrending, and CrAO datasets, as well as publicly available photometry from the ASAS-SN and AAVSO surveys. 
    The K2 light curve was rescaled to the 7 CrAO measurements taken contemporaneously. 
    The normalized K2 fluxes were thus converted to magnitudes applying a 12.3 mag zero-point offset. Over the time frame where they overlap, the amplitude of the K2 light curve appears somewhat smaller than that of the V-band CrAO light curve, presumably reflecting the longer effective wavelength of the K2 wide-band filter.
    Interestingly, the low-frequency part of the K2 light curve, most notably the brightness ``bump'' seen over its first part, is recovered over the slightly overlapping parts of the CrAO and ASAS-SN light curves, which indicates intrinsic longer term variations superimposed onto the spot modulation. 
    
    \begin{figure*}
        \centering
        \includegraphics[width=.45\textwidth]{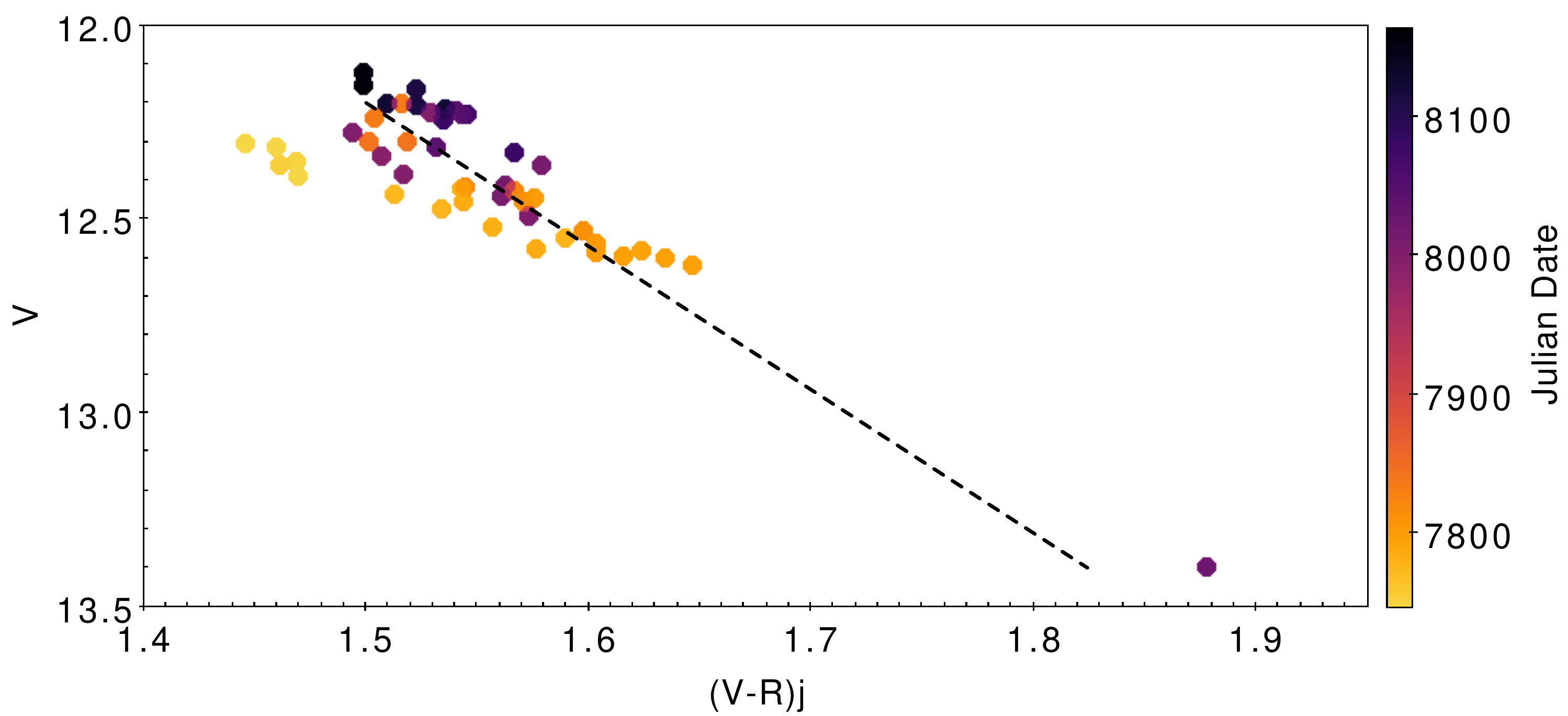}
        \includegraphics[width=.45\textwidth]{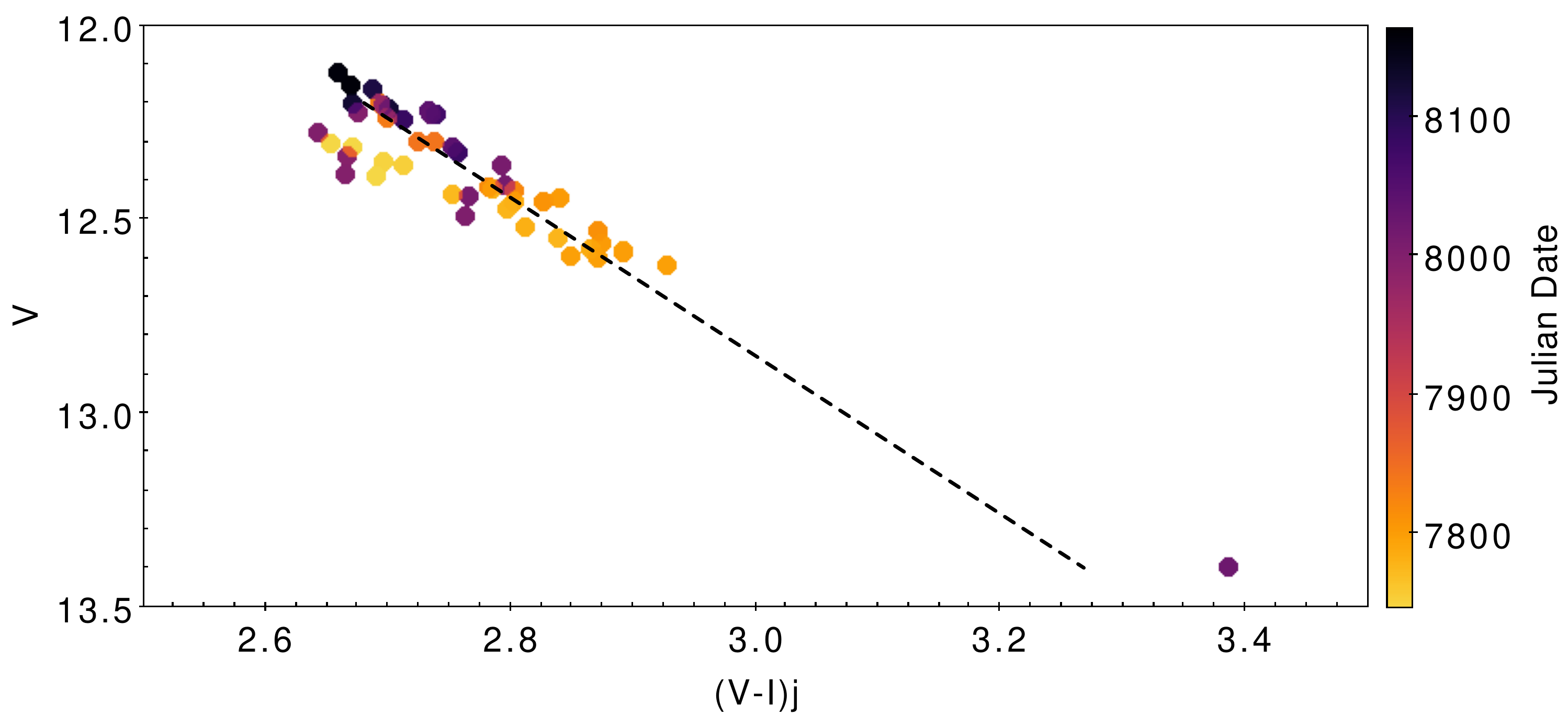}
        \caption{Color-magnitude plots from CrAO VR$_J$I$_J$ photometry. {\it Left:} (V-R)$_J$ vs. V. {\it Right:} (V-I)$_J$ vs. V. The source gets redder when fainter. The dashed line is the expected reddening line for ISM extinction. The photometric measurements are color-coded by Julian date to distinguish the 2 observing seasons: lighter points correspond to the first epoch, darker points to the second one (see Fig.~\ref{fig:lc}). We note that the deep faintening event of about 1.2 mag in the V-band is accompanied by a strong reddening of the system.}
        \label{fig:color}
\end{figure*}

   Photometric data points were also available from the ASAS-SN survey \citep{Jayasinghe19}. 
   HQ Tau was observed in both the V-band and g-band filters during this period. 
   Taking advantage of the overlap between the V- and g-band light curves, the g-band measurements were arbitrarily shifted by -0.8 mag in order to match the V-band ones. 
   The resulting ASAS-SN light curve contains a few measurements during the first half of the ESPaDOnS observations and many additional measurements before and after the run was completed. 
   It thus provides an estimate of the amount of variability the source exhibited at the time of the spectroscopic observations, performed about 6 months after the K2 monitoring. 
   Overall, it suggests relatively mild variability at this epoch, amounting to about 0.2 mag in amplitude in the V-band. 
   A significant dimming event, which lasted for about a month just prior to the start of the spectroscopic observations had apparently ceased by the time of the ESPaDOnS measurements. 
   We notice, however, that the mean flux level was changing over the spectroscopic run, with the source being about 0.3 mag brighter toward the end of the run. 
   Both the amplitude of variability and the low-frequency variations are not unlike those seen in the spot-driven K2 light curve obtained several months earlier. 
   This suggests that at the time of the ESPaDOnS observations, the source was in a state of relatively mild periodic variability, and probably not in a drastic dipper state as reported at some other epochs by \cite{Rodriguez17}, in spite of the occurence of a recent dimming event. 
   
   Indeed, we verified that the ASAS-SN g-band and V-band datasets restricted to the 7 days following the ESPaDOnS run (JD 2,458,067.8-2,458,073.7) exhibit a smoothly varying light curve consistent with a period of 2.424 days when folded in phase. Using the same epoch for the origin of phase as for the K2 light curve above (i.e., JD 2,457,823.81), we find that the photometric minimum of the ASAS-SN dataset occurs around phase $\sim$0.56, which we estimated by interpolating  
   the two lowest photometric measurements occuring at phase 0.47 and 0.66, respectively. 
    Hence, the ASAS-SN photometric minimum is not far from phase 0.50 of the K2 photometric minimum, the slight difference being easily accounted for by the 0.002~d uncertainty on the K2 period.
    The near conservation of phase over the 155 day-long temporal gap stretching between the end of the K2 observations and the beginning of the ESPaDOnS run suggests the modulation results from a long lived, relatively stable spot structure.  For the rest of the paper, we thus use the following ephemeris:
        \begin{equation}
            HJD(d) = 2,457,823.81 + 2.424E,
        \end{equation} 
    where $E$ is the rotational phase of the system.

   Color variations associated to the brightness changes are shown in Fig.~\ref{fig:color}. 
   In both the (V-R)j and (V-I)j colors, the source becomes redder when fainter. 
   The color slope of the small scale variations likely results from spot modulation \citep{Herbst94, Venuti15}. 
   A single deep faintening event was recorded on JD 2,458,025.6 in the VR$_J$I$_J$ filters of the CrAO dataset, and is confirmed by the single-filter ASAS-SN light curve (see Fig.~\ref{fig:lc}). 
   The color plot shows that as the brightness decreased by about 1.2 mag in the V-band, the system became much redder, with a color slope close to that expected for extinction by ISM-like grains.  
   This suggests that the dimming event was caused by circumstellar dust crossing the line of sight.


\subsection{Stellar parameters}
\label{subsec:stellarparam}
	
\begin{figure}
  \centering
  \includegraphics[width=0.5\textwidth]{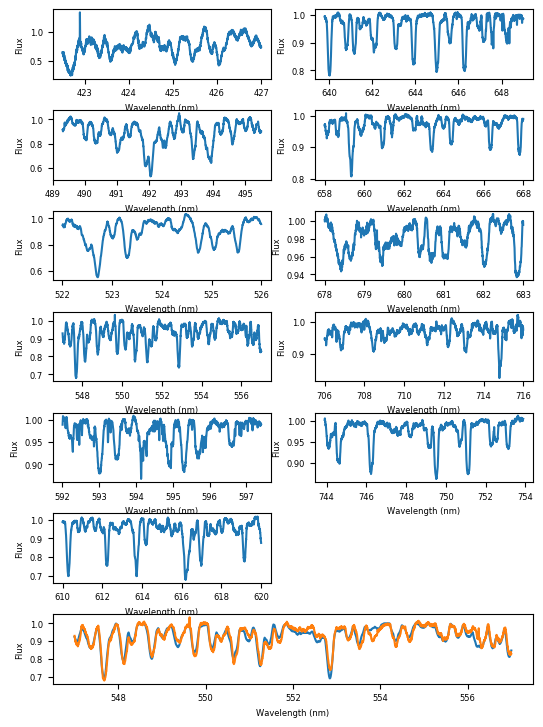}
  \caption{{\it Upper six rows:} Spectral windows selected from HQ Tau's mean ESPaDOnS spectrum to derive stellar parameters. {\it Bottom row:} As an example, HQ Tau's mean spectrum (blue) fitted with a ZEEMAN synthetic spectrum (orange) in the 547-557~nm window. }
  \label{fig:lma}
\end{figure}

We used the ESPaDOnS high resolution spectra to derive HQ Tau's stellar parameters, namely effective temperature (\teff), and the radial (\vrad), rotational (\vsini), and microturbulent (\vmic) velocities.
    We averaged the 14 ESPaDOnS spectra gathered during the campaign and fit synthetic spectra calculated with the ZEEMAN code \citep{Landstreet88,Wade01,Folsom12} based on MARCS stellar atmosphere grids \citep{Gustafsson08}, VALD line lists \citep{Ryabchikova15}, and including the same oscillator strength corrections as those used in \cite{Folsom16}.
    We explored a range of \teff, \vsini, \vrad, and \vmic, and obtained the best fit to HQ Tau's mean spectrum through a $\chi ^2$ minimization procedure using a Levenberg-Marquardt algorithm \citep[see][]{Folsom13}.
    The fit was performed on 11 independent spectral windows spanning the range from 422 to 754 nm, each with a width ranging from 4 to 10~nm. Each window contains well resolved, relatively unblended photospheric lines, and is devoid of emission features.
    The spectral windows are shown in Fig.~\ref{fig:lma}.

    In order to derive the stellar properties from spectral fitting, we first fixed the macroturbulent velocity to 2~\kms, \logg\ to 4.0, and assumed solar metallicity, in other words, values that are typical of low-mass PMS stars in the solar neighborhood \citep{Padgett96, James06, Santos08, Taguchi09, Dorazi11}. 
    An example of the resulting fit is shown in Fig. \ref{fig:lma}.
    Once stellar parameters were obtained for each spectral window, we averaged them and computed the rms dispersion, and then removed the windows that yielded results beyond 1$\sigma$ from the mean value. 
    Usually, between 3 or 4 windows were thus rejected. We thus derive \teff\ = 4997$\pm$160~K, 
   \vsini\ = 53.9 $\pm$ 0.9~\kms, \vrad\ = 6.64 $\pm$ 0.71~\kms, and \vmic\ = 1.4 $\pm$ 0.2~\kms.  The \teff\ of 4997~K we derive is consistent with a spectral type K0-K1 according to \cite{Pecaut13} conversion scale.

We next obtained an estimate of HQ Tau's luminosity, based on its 2MASS J-band magnitude \citep[J=8.665$\pm$0.024,][]{Skrutskie06}, using the Gaia parallax  \citep[$\pi=6.304 \pm 0.213$ mas,][]{Gaia18}. Extinction was computed from the (V-J) color excess, using intrinsic colors from \cite{Pecaut13}, which yields A$_V$=2.60 $\pm$ 0.1~mag and A$_J$=0.73 $\pm$ 0.03~mag, consistent with the value found by \cite{Herczeg14}. We applied bolometric corrections from \cite{Pecaut13}, and obtained 
 \lstar = 3.91 $\pm$ 0.65~\lsun, and \rstar = 2.64 ~$\pm$ 0.31~\rsun.  
 	
\begin{figure}
        \centering
        \includegraphics[width=0.5\textwidth]{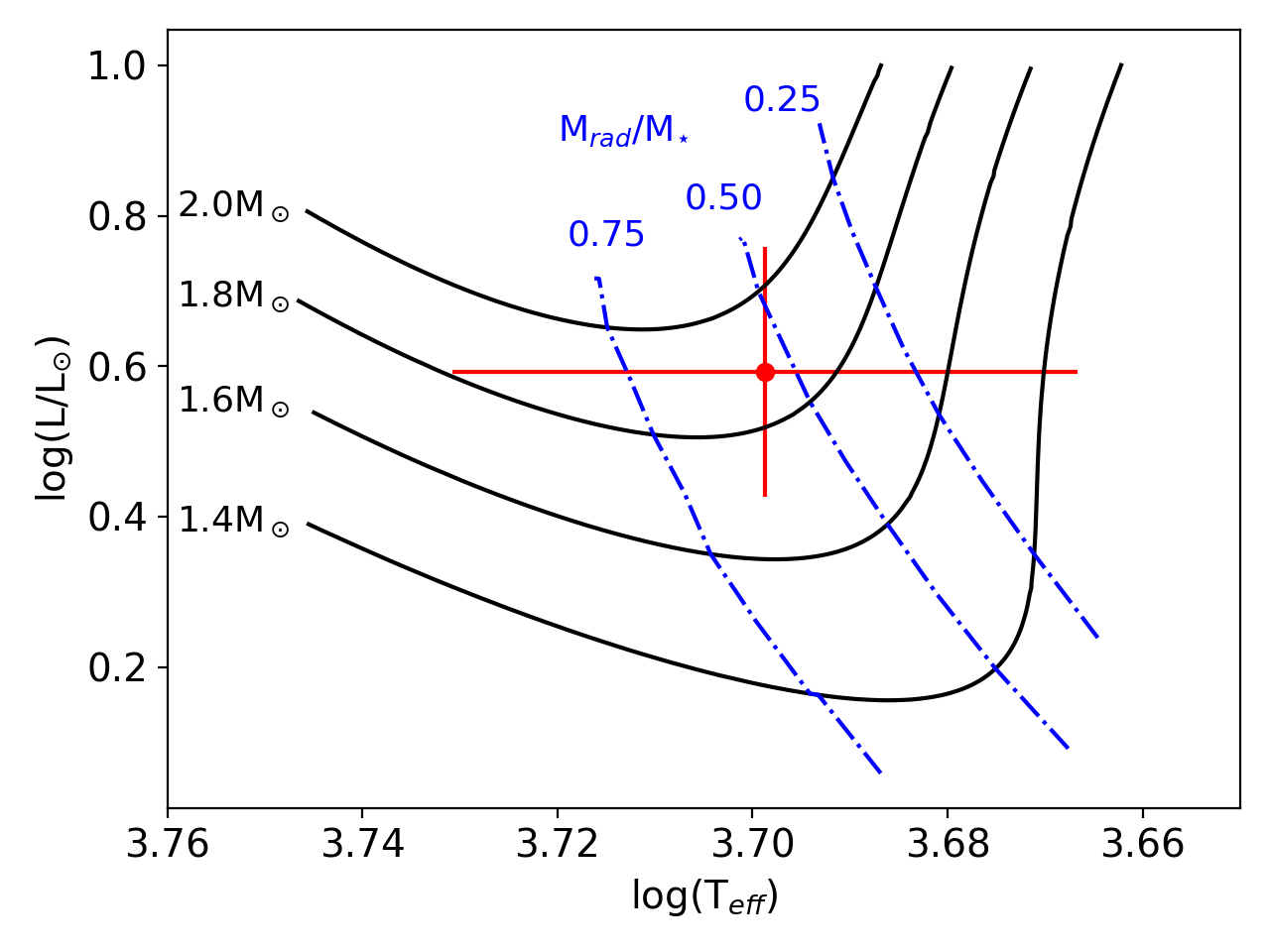}
        \caption{HQ Tau's position in the HR diagram. The red cross illustrates HQ Tau's position with corresponding uncertainties. The black curves are the evolutionary tracks from CESTAM PMS models, with the  corresponding mass indicated on the left. The blue curves show the position where the radiative core of the star reaches 75\%, 50\% and 25\% of the stellar mass.}
        \label{fig:hrd}
\end{figure}

	We then plotted HQ Tau in an Hertzsprung-Russell (HR) diagram and used a grid of CESTAM evolutionary models \citep{Marques13, Villebrun19} to derive its mass and internal structure (see Fig.~\ref{fig:hrd}). From bilinear interpolation in model grids, we obtain \mstar = 1.87 $^{+0.21}_{-0.55}$ \msun, and an age of $\sim$2~Myr. According to evolutionary models,  the star, is partly radiative with \mrad = 0.55 $^{+0.35}_{-0.55}$~\mstar and \rrad = 0.51$^{+0.14}_{-0.51}$~\rstar, where \mrad\ and \rrad\ are the mass and radius of the radiative core, respectively.
	
	 We find that HQ Tau is a fast rotator with \vsini = 53.9~\kms. Combining the stellar radius with the star's \vsini\ and the K2 rotational period, we derive the inclination angle of the rotational axis onto the line of sight, $i$=75$^{+15}_{-17}$ deg, indicative of a highly inclined system.

\subsection{Spectroscopic properties and variability}
\label{subsec:spectro}

    We investigate the spectral variability of the system by analyzing the 14 ESPaDOnS spectra taken over 12 nights.
    Table \ref{tab:rad_vel} summarizes radial velocity and line equivalent width measurements. 
    
\subsubsection{Radial velocity variations}

\begin{figure*}[t]
\centering
  \includegraphics[width=0.8\textwidth]{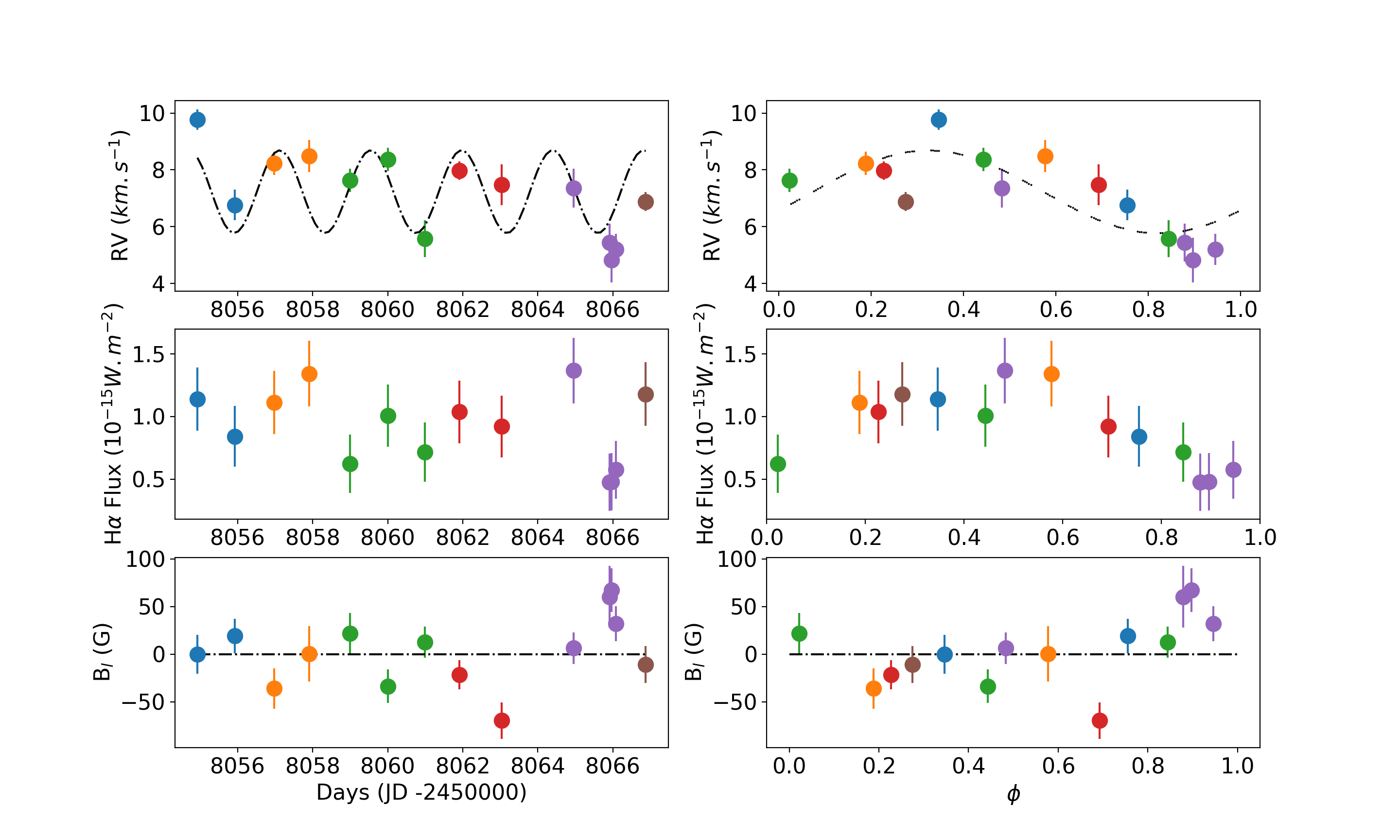}
  \caption{{\it From top to bottom:} Radial velocity, \ha\ line flux, and average longitudinal magnetic field as a function of Julian date {\it (left)} and rotational phase {\it (right)}. The sinusoidal fit, forced at the K2  photometric period, to radial velocity variations is shown as a black dash-dotted curve in the top panels. The B$_l$=0~G line is drawn for reference in the bottom panels.}
  \label{fig:param-mod}
\end{figure*}

    Radial velocity measurements were obtained by 
    cross-correlating each HQ Tau's spectrum with the spectrum of a spectral template. 
    We used the ESPaDOnS observations of the Hyades cluster member Melotte 25-151 as a template, a slowly rotating (\vsini\ = 4.8 \kms) K2-type star with \vrad = 37.98 \kms\ and \teff = 4920~K \citep{Folsom18}, reduced and normalized in the same way as HQ Tau.
     We computed the cross correlation function (CCF) over 5 spectral windows (542-547, 558-563, 585-587, 608-613, and 639-649 nm), and  
    fit a Gaussian profile to derive the radial velocity difference between HQ Tau and the template. We averaged the results over the 5 spectral windows to get a mean value of \vrad\ and its rms uncertainty. 
	The results are listed in Table \ref{tab:rad_vel}
	and the radial velocity curve is shown in Fig.~\ref{fig:param-mod}.    
	
	 The radial velocity appears to be modulated and a sinusoidal fit yields a period of $2.48 \pm 0.16$ d using the 1$\sigma$ confidence level on $\chi^2$ minimization,  consistent with the stellar rotation period within uncertainties. We therefore ascribe this modulation to surface spots. The phased radial velocity curve is shown in Fig. \ref{fig:param-mod}. Indeed, the sinusoidal fit indicates that the mean radial velocity (<\vrad>=7.22 $\pm$ 0.27~\kms) occurs around phase 0.6, which is expected  when the spot faces the observer \citep{Vogt83} and is consistent with the photometric minimum of the contemporaneous ASAS-SN light curve.
     The amplitude of the \vrad\  modulation amounts to hardly a tenth of the star's vsini, which suggests it is mostly driven by a large area and high latitude cold spot. We also notice a regular downward drift of \vrad\ with an amplitude of 2-3~\kms\ over a timescale of 10 days. We will come back to this feature in Section 4.

  \begin{table}	
  \small
	    \centering
	    \begin{tabular}{l|ll|llll}
	        \hline
	        &&&\multicolumn{4}{| c }{EW (\AA)}\\
	        HJD & \vrad & $\sigma$\vrad & \ha & \hb & CaII & CaII \\
	        (2450000+) & \multicolumn{2}{ c }{(\kms)}  &   &  & $\lambda$8542 & $\lambda$8662 \\
	        \hline
	        	        8054.93481 & 9.77 & 0.36 & 3.25 & 0.96 & 1.66 & 1.42 \\
	        8055.92519 & 6.76 & 0.53 & 2.40 & 0.09 & 1.16 & 0.96 \\
	        8056.97501 & 8.23 & 0.40 & 3.18 & 0.96 & 1.59 & 1.36 \\
	        8057.91808 & 8.48 & 0.57 & 3.84 & 1.30 & 1.80 & 1.63 \\
	        8058.99693 & 7.63 & 0.41 & 1.77 & 0.06 & 1.09 & 0.94 \\
	        8060.01605 & 8.37 & 0.41 & 2.88 & 0.88 & 1.61 & 1.46 \\
	        8060.98848 & 5.58 & 0.65 & 2.05 & 0.12 & 1.25 & 1.17 \\
	        8061.91652 & 7.97 & 0.33 & 2.96 & 0.84 & 1.52 & 1.46 \\
	        8063.04627 & 7.47 & 0.72 & 2.64 & 0.59 & 1.46 & 1.31 \\
	        8064.96108 & 7.35 & 0.69 & 3.91 & 1.03 & 1.88 & 1.65 \\
	        8065.92044 & 5.44 & 0.67 & 1.36 & 0.28 & 0.92 & 0.82 \\
	        8065.96526 & 4.82 & 0.78 & 1.37 & 0.29 & 0.92 & 0.91 \\
	        8066.08352 & 5.20 & 0.54 & 1.64 & 0.06 & 1.13 & 1.05 \\
	        8066.88036 & 6.88 & 0.34 & 3.37 & 1.16 & 1.60 & 1.46 \\
	        \hline
	    \end{tabular}
	    \caption{Photospheric radial velocity and equivalent widths of selected emission lines.}
	    \label{tab:rad_vel}
	\end{table}
  
\subsubsection{Emission line profiles}
\label{subsub:lineprof}

\begin{figure*}
    \centering
    \includegraphics[width=0.9\textwidth]{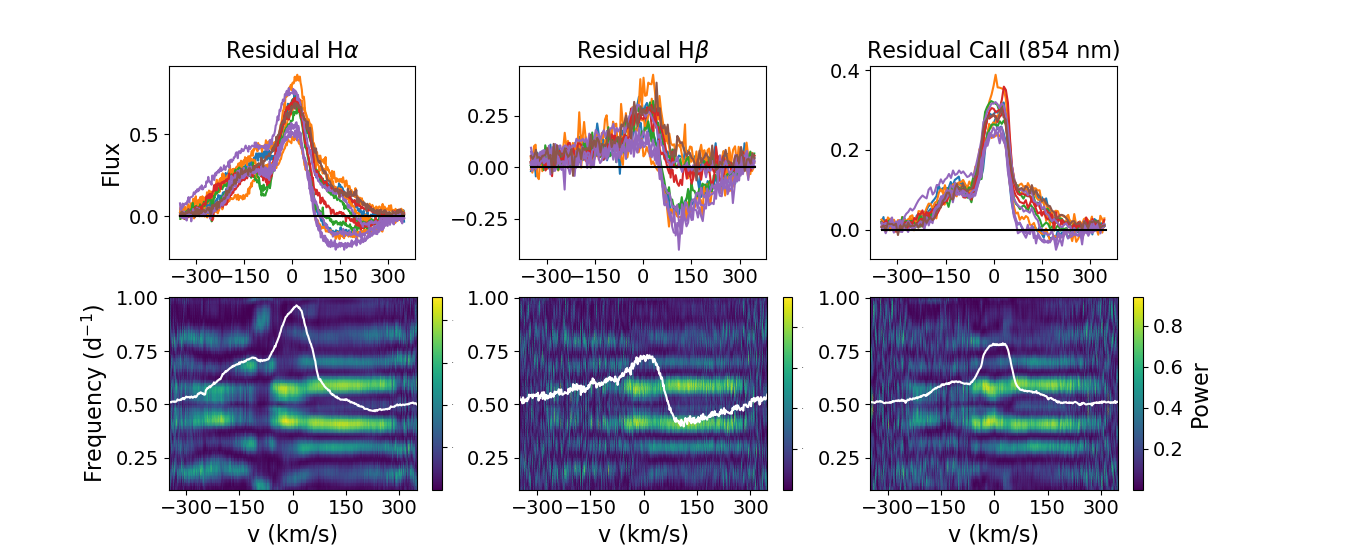}
    \caption{{\it Top:} Residual \ha\ (left), \hb\ (middle), and CaII 854.2 nm (right) emission lines of HQ Tau spectra.  {\it Bottom:} Corresponding  2D periodograms with the mean profile superimposed as a white curve. The maximum power of the periodogram is in yellow and the minimum in dark blue. The red and blue wings of \ha\ and the red wing of \hb\ and CaII, as well as the center of the 3 line profiles, exhibit a periodic modulation at $f \approx 0.4$  d$^{-1}$, corresponding to the rotational period of the star. The power seen at $f \approx 0.6$ d$^{-1}$ is the 1-day alias induced by the observational window.}
    \label{fig:ha_hb_ca}
\end{figure*}

We computed residual emission line profiles by subtracting the rotationally broadened photospheric profiles of the template Melotte 25-151 from HQ Tau's profiles.  
    \ha, \hb, and the CaII infrared triplet (IRT) are the only lines in HQ Tau's spectrum exhibiting significant emission flux.  
    The 3 lines of the CaII IRT are similar in shape but the line at 866.2 nm may be affected by a hydrogen line of the Paschen series and the line at  849.8 nm exhibited a variation far lower than the other two. 
    We therefore focused on the 854.2 nm component.
    Figure \ref{fig:ha_hb_ca} shows the \ha, \hb, and the selected CaII IRT residual line profiles. The raw line profiles are shown in Fig.~\ref{fig:temp}.
\begin{figure*}
  \centering
  \includegraphics[width=0.8\textwidth]{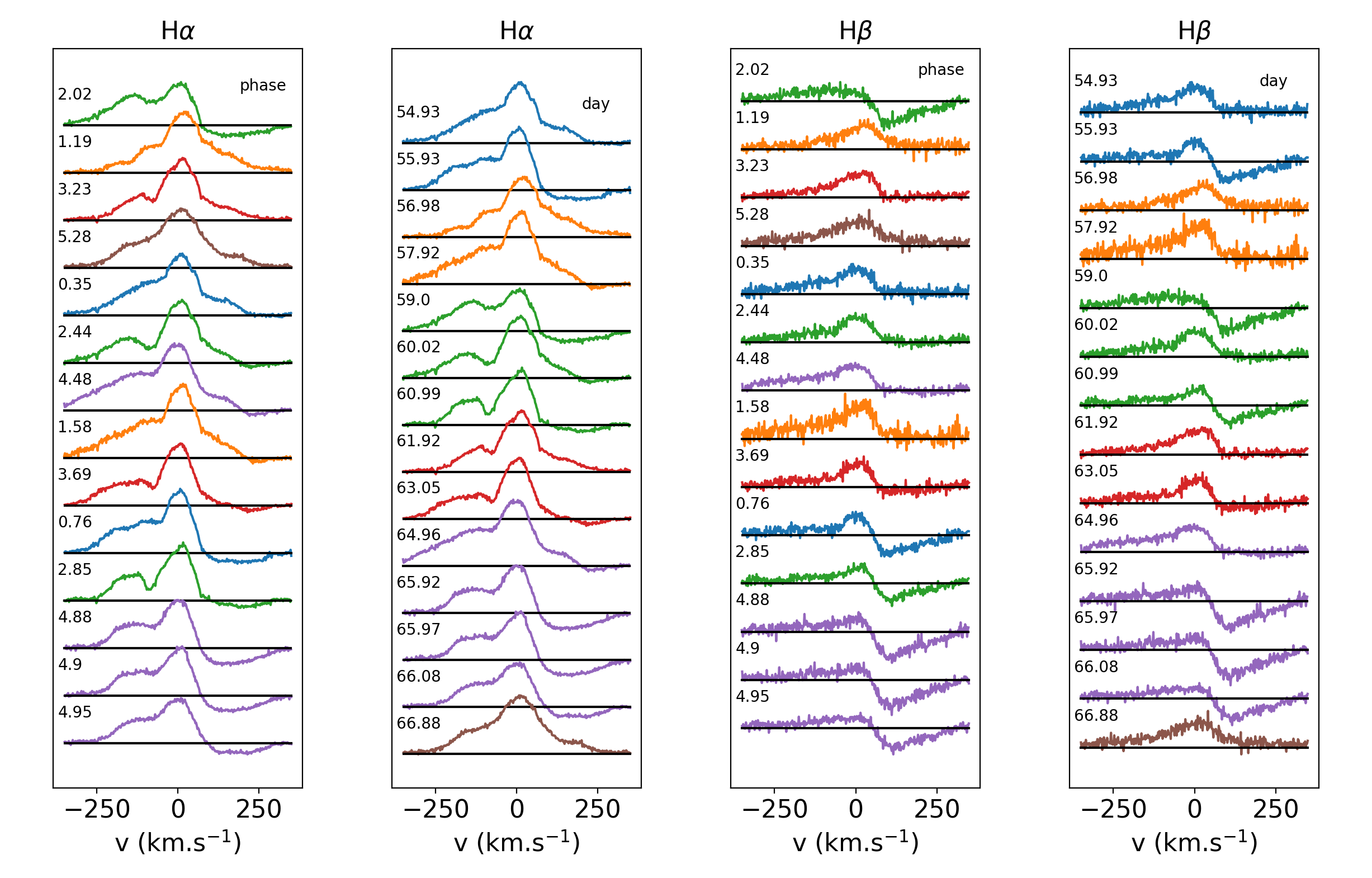}
  \caption{Residual \ha\ (left panels) and \hb\ (right panels) profiles sorted by rotational phase and by day of observation. The successive rotational cycles are shown by different colors, as in Fig.~\ref{fig:param-mod}. }
  \label{fig:hbeta}
\end{figure*}

    The H$\alpha$ line profile exhibits high velocity wings to the blue and red sides, up to about 300 \kms. The red wing displays large variability, with at times signs of high velocity redshifted absorptions reaching below the continuum, which are Inverse P Cygni (IPC) profiles. 
    The blue wing also displays significant variability. 
    The H$\beta$ line is dominated by deep IPC signatures extending up to about 300 \kms\ seen in nearly half of the observations.
    The CaII line also exhibits strong variability from the line center to its red wing, with however only marginal IPC components below the continuum level.
    
    We computed periodograms in each velocity bin across the line profiles, the result of that is a 2D periodogram for each line, which is shown in Fig.\ref{fig:ha_hb_ca}.    In all 3 lines, a clear periodicity is seen at a frequency of 0.4 d$^{-1}$ (P=2.5 d) with a FAP reaching $\sim$ 10$^{-4}$ and a typical value of $\sim$ 10$^{-2}$, corresponding to the stellar rotational period, extending from the line's central peak all the way to the red wing.  This is a clear indication that the IPC components are modulated by stellar rotation, as expected for funnel flow accretion.   
    We do notice significant power as well at about the same frequency at highly blueshifted velocities ($\sim$-200~\kms) in the H$\alpha$ profile, which might also be present in the CaII line profile.
    The peak at a frequency of 0.6 d$^{-1}$ is the 1-day alias of the 0.4 d$^{-1}$ frequency. It appears clearly here, though at a higher FAP of 0.1, due to the night-to-night sampling of ESPaDOnS spectra.

    Figure \ref{fig:hbeta} shows the residual \ha\  and \hb\ profiles ordered by day and by phase. The phase ordering illustrates well the periodicity of IPC components: they appear over the 5  rotational cycles covered by the spectral series at specific phases, from 0.69 to 0.02 in both profiles, 
    with a maximum depth around phase 0.90. 
    The depth of the IPC components appears to slightly vary from one rotational cycle to the next, being stronger at phase 4.90 than at phase 2.85 for instance.

\begin{figure}
    \centering
     \begin{subfigure}{0.20\textwidth}
        \centering
        \includegraphics[width=\textwidth]{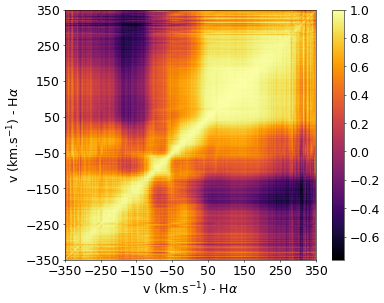}
        \caption{\ha\ vs. \ha}
        \label{fig:cm_haha}
    \end{subfigure}
     \begin{subfigure}{0.20\textwidth}
        \centering
        \includegraphics[width=\textwidth]{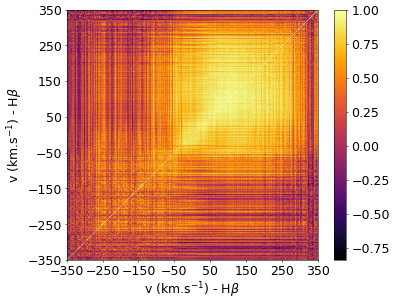}
        \caption{\hb\ vs. \hb}
        \label{fig:cm_hbhb}
    \end{subfigure}
    
     \begin{subfigure}{0.20\textwidth}
        \centering
        \includegraphics[width=\textwidth]{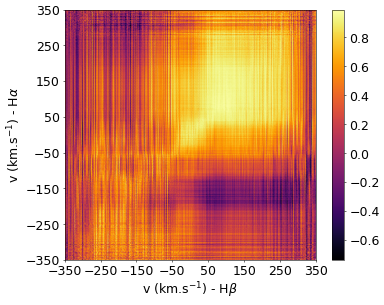}
                \caption{\hb\ vs. \ha}
        \label{fig:cm_hbha}
    \end{subfigure}
    \begin{subfigure}{0.20\textwidth}
        \centering
        \includegraphics[width=\textwidth]{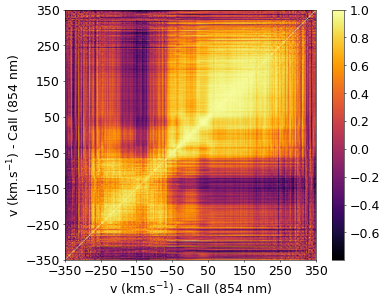}
                \caption{CaII vs. CaII}
        \label{fig:cm_ca854_ca854}
    \end{subfigure}
    
    \begin{subfigure}{0.20\textwidth}
        \centering
        \includegraphics[width=\textwidth]{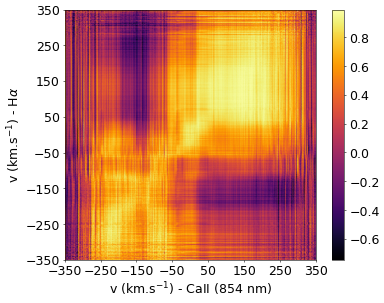}
                \caption{\ha\ vs. CaII}
        \label{fig:cm_ca854_Ha}
    \end{subfigure}
    \begin{subfigure}{0.20\textwidth}
        \centering
        \includegraphics[width=\textwidth]{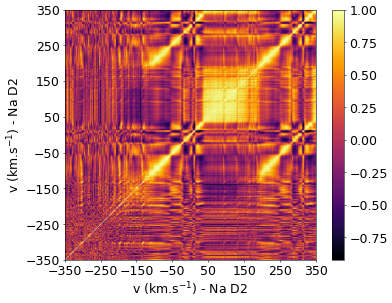}
                \caption{NaD\ vs. NaD}
        \label{fig:cm_na_na}
    \end{subfigure}
    
    \caption{Correlation matrices of the line profiles.}
    \label{fig:cm}
    \end{figure}

   We investigate the relationship between the various components seen in the line profiles by computing correlation matrices \citep{Johns95b, Oliveira00, Alencar02, Kurosawa05}. Correlation matrices consist of the computation of Pearson's linear correlation coefficient on line intensity between 2 velocity channels of the same or different line profiles. 
    The coefficient approaches 1 for a strong correlation, 0 for no correlation or -1 for anticorrelated intensity variations.
    Correlation matrices are shown in Fig. \ref{fig:cm}.
    
A strong correlation is seen between velocity channels within the red wing of all profiles, and within the blue wing  for \ha\ and CaII. These correlations appear as bright squares in the correlation matrices, located on each side of the line center and extending to high velocities. As discussed above, the strong correlation within the red side of the line profiles is probably related to the periodic appearance of IPC components, which extends from close to the line center up to +300~\kms. Strikingly, little correlation is seen between the red and blue wings of the line profiles, suggesting that their variability is driven by unrelated physical processes, presumably accretion for redshifted velocity channels and outflows for blueshifted ones. Yet, an interesting anti-correlation appears between a restricted range of blueshifted velocity channels, from about -130 to -220 \kms, and the redshifted part of the line profile. This is clearly seen in the \ha\ correlation matrix and also appears in the CaII matrix. This indicates that, as the IPC components appear on the red side of the line profile, the high velocity part of the blue wing becomes more intense, perhaps the signature of an accretion-driven high velocity outflow \citep[e.g.,][]{Johns95a}.

    \subsubsection{The NaI D profile }

\begin{figure}
        \centering
        \includegraphics[width=.50\textwidth]{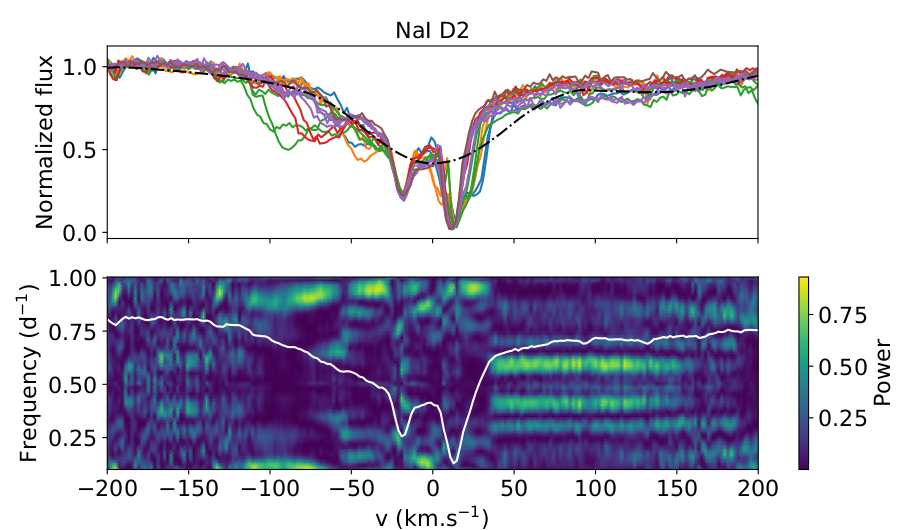}
        \caption{{\it Top:} The NaI D2 line appearing in the 14 spectra of HQ Tau. The black dash-dotted line is a synthetic NaI D2 line produced using the ZEEMAN code with HQ Tau's parameters. {\it Bottom:} 2D periodogram of the NaI D2 line.}
        \label{fig:na_sup}
    \end{figure}

The NaI doublet is sensitive to wind signatures \citep{Mundt84}, and can thus help to better understand the accretion-ejection connection. The NaI D lines are seen in absorption in HQ Tau's spectrum and exhibit significant variability (see Fig.~\ref{fig:na_sup}). As the two lines of the doublet have similar profiles and behavior, we focus here on the 589.0 nm line (NaI D2), leaving aside the 589.6 nm twin line, which lies at the edge of a spectral order. 

Two prominent, relatively narrow, and apparently stable absorption components are seen around the line center. The stable component located at +10 \kms\ (V$_{helio}$$\simeq$ 17~\kms) was previously reported by \cite{Pascucci15} and ascribed to local interstellar absorption. This narrow and stable absorption component is also present at the same redshifted velocity in all our spectra in the KI 770 nm line profile. We notice however significant variability on both the blue and red sides of this component in the NaI D2 line, over a range of about $\pm$10~\kms, which indicates an additional source of variable absorption, linked to the stellar system itself. These transient components are also seen in the KI profile. The second narrow absorption component is blueshifted and located at -20~\kms (V$_{helio}$$\simeq$ -13~\kms). This component is also apparent in the NaI D line profile reported in \cite{Pascucci15}, but is not discussed there. CI Tau, which is located only 0.5 degrees away from HQ Tau, does not exhibit this second component, while it does display the redshifted one. This suggests the blueshifted narrow absorption component is not due to interstellar cloud absorption. Yet, its narrow width compared to photospheric lines and its stability suggest it is not related to the inner variable system. This component could conceivably be the signature of the distant, cold part of a stellar wind, as discussed in \cite{Mundt84}.

    \begin{figure}
        \centering
        \includegraphics[width=.45\textwidth]{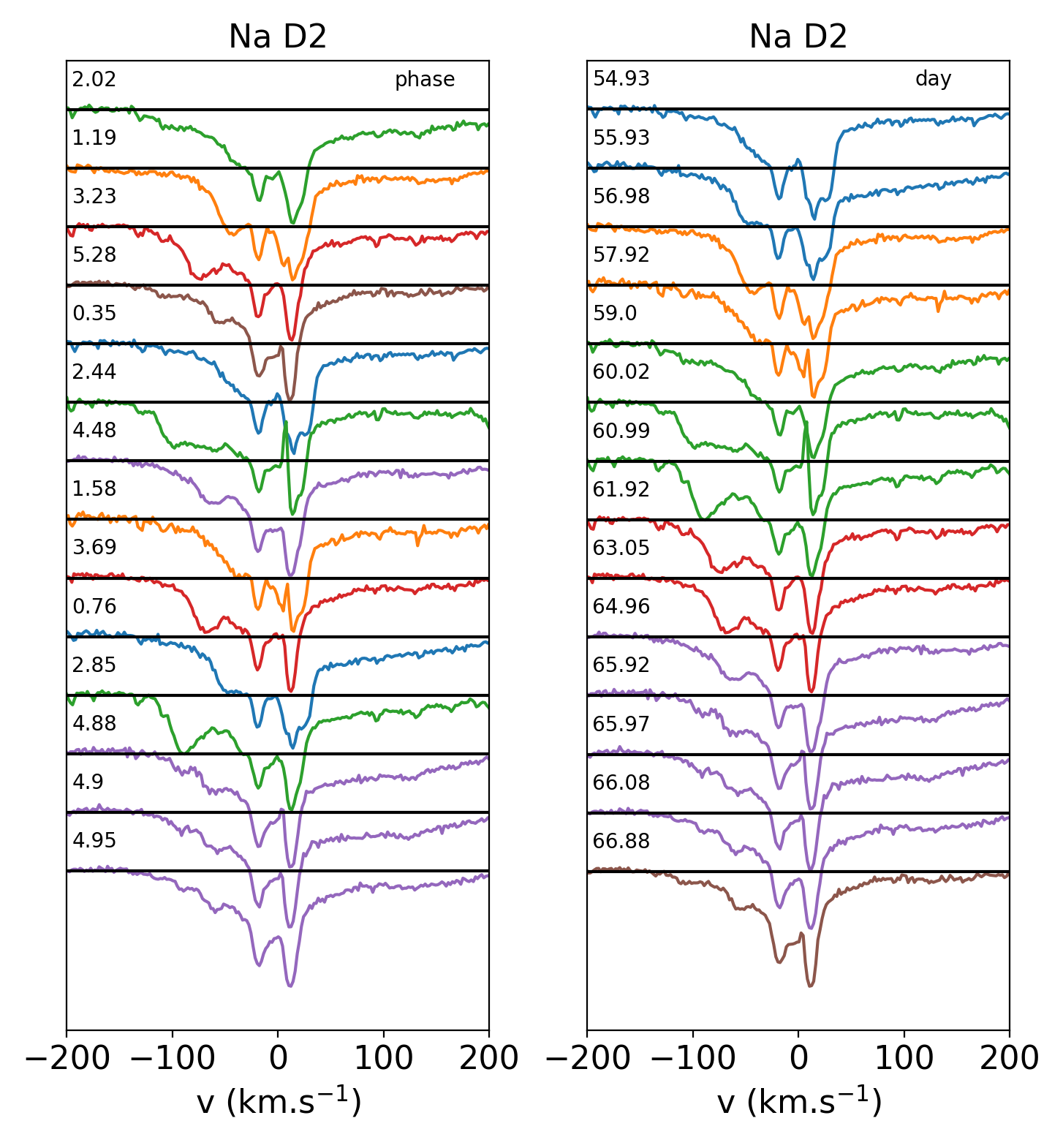}
        \caption{NaI D2 line profiles sorted by rotational phase {\it (left)} and by day of observation {\it (right)}. The successive rotational cycles are shown by different colors, as in Fig.~\ref{fig:param-mod}.}
        \label{fig:prof_na}
    \end{figure}   

The variability seen across the NaD line profile is relatively complex, with different components. The 2D periodogram shown in Fig.~\ref{fig:na_sup} reveals that the red side of the profile, from about +40 to +150 \kms, varies periodically, at the stellar rotation period. The comparison of successive NaI D2 profiles ordered in phase, shown in Fig.~\ref{fig:prof_na}, suggests this is due to the appearance of IPC components seen between phase 0.69 and 0.02, as reported above for Balmer and CaII emission lines. In contrast, the variability seen in the blue wing of the line profile is not periodic. Sporadic blueshifted absorption components appear in nearly half of the observations, over velocity channels ranging from about -120 to -30 \kms. Figure~\ref{fig:prof_na} shows the occurrence of deep, high velocity blueshifted absorption components from JD 2,458,060 to 2,458,066 covering velocity channels from about -120 to -50 \kms, with evidence for a gradual shift in velocity over this timescale.  Lower velocity absorption components are also seen for a couple of days from JD 2,458,056 to 2,458,057 around -60 to -30 \kms. 
    None of these blueshifted components appear to last over the whole observing period. 
  They appear to be transient phenomena occurring only during specific cycles, are of limited duration, and are not modulated by stellar rotation. These are presumably the signatures of episodic outflows.    
     
Finally, Fig.~\ref{fig:cm_na_na} presents the correlation matrix of the Na D2 line. It resembles the matrices of the other emission lines in the red part of the profile, being similarly modulated by the IPC components. It does not show any correlated variability in the blue wing, as none is expected from the sporadic, independent outflows components analyzed above.

\subsubsection{Mass accretion rate}

    We computed the mass accretion rate using the relationship between the accretion luminosity and the luminosity in the H$\alpha$ residual line \citep{Alcala17}.
    \cite{Duchene17} notice that the H$\alpha$ emission line of HQ Tau might not be a good accretion indicator, due to its weakness, though its width and double peaked shape favor an accretion origin.
    In order to confirm the mass accretion rate values found with H$\alpha$, we also computed the mass accretion rate using the H$\beta$, and CaII IRT residual lines.
	The line flux was computed from the line equivalent width as 
        $F_{line} = F_0 \cdot EW \cdot 10^{-0.4m_{\lambda}}$, 
    where $F_{line}$ is the line flux, $F_0$ is the reference flux in the selected filter, EW the line equivalent width, and $m_{\lambda}$ is the extinction-corrected magnitude of HQ Tau in the selected filter, namely R$_C$ for H$\alpha$, R$_J$ for the Ca II IRT, and B for H$\beta$. 
    The line luminosity is then derived from 
        $L_{line} = 4 \pi d^2 F_{line}$,
    where $d$ is the distance to HQ Tau. The accretion luminosity follows from the line luminosity by 
       $ \log (L_{acc}) = a\log (L_{line})+b$,
    where $a$ and $b$ are coefficients taken from \cite{Alcala17}. 
    The mass accretion rate is obtained from:
 \begin{equation}
        L_{acc} = \frac{G M_{\star}\dot{M}_{acc}}{R_{\star}}\bigg[1-\frac{R_{\star}}{R_{t}}\bigg],
        \label{eq:macc}
    \end{equation}
    where $R_{t}$ is the magnetospheric truncation radius, typically $5R_{\star}$ for cTTs \citep{Bouvier07b}. 
        
    As the equivalent width of Balmer lines is strongly affected by the IPC components, we computed the mean value of \lacc\ and \macc\ by averaging the results excluding the observations showing strong IPC profiles. The mass accretion rates we deduce from the various lines are consistent within 2$\sigma$.
    This yields \lacc=0.019 $\pm$ 0.005 \lsun\ and \macc = 1.26 $\pm$ 0.35 10$^{-9}$ \msunyr\ for the system.  
    As \ewha\ is relatively weak in HQ Tau, the emission line flux might include a non negligible chromospheric contribution. In order to quantify it, we considered a sample of wTTSs with similar spectral type and \vsini\ and (or) rotation period as HQ Tau, which were chosen from the list of  \cite{James06}.
    We thus derived <\ewha> = 0.7$\pm$0.1 ~\AA\ for these wTTSs, assumed this value to be the chromospheric contribution, and subtracted it from HQ Tau's \ewha. We then recomputed \macc\ from the different line estimates to obtain \macc = 1.16 $\pm$ 0.35 10$^{-9}$ \msunyr.
    The temporal variation of the \ha\ line flux, a proxy for \macc, is shown in Fig. \ref{fig:param-mod}. It is clearly modulated and a  
    sinusoidal fit yields a period of 2.4 d, consistent with the stellar rotational period. The peak-to-peak amplitude of the modulation is about a factor 3, typical of cTTs \macc\ variability on these timescales \citep{Venuti14}. 
        
    Another signature of accretion is the so-called veiling.
    Veiling is an additional continuum component emitted by the accretion shock, which fills the photospheric lines.
    In order to compute the amount of veiling in HQ Tau's spectrum, we used the non-accreting star Melotte 25-151 as a template. Varying the amount of continuum line filling over the spectral window 639-649~nm in the template spectrum, we matched HQ Tau's spectrum by a $\chi^2$ minimizing method: 
    
    \begin{equation}
        I(\lambda) = \frac{I_t(\lambda)(1+r)}{1+I_t(\lambda)r},
        \label{eq:chi-mod}
    \end{equation}
    where $I_t$ is the intensity spectrum of Melotte 25-151, the template, and $r$ the fractional veiling that is the  excess continuum flux divided by the stellar continuum flux at a given wavelength. We derived a weak veiling at 640~nm ranging from 0.15 to 0.20 in the 14 HQ Tau spectra, as expected for the modest mass accretion rate and bright photosphere of the system, with uncertainties of order of 0.20, thus preventing us from detecting any significant temporal variations.


\subsection{Spectropolarimetry}

\begin{figure*}[t]
    \centering
    \includegraphics[width=.99\textwidth]{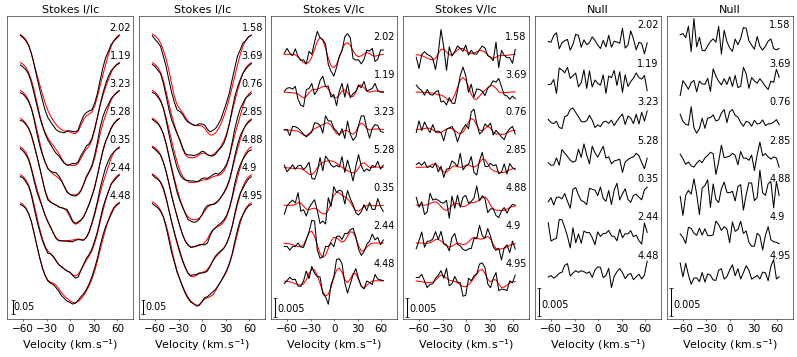}
    \caption{HQ Tau's Stokes I {\it (left)}, V {\it (middle)}, and Null {\it (right)} profiles (black) ordered by increasing fractional phase,  and the fit produced by the ZDI analysis (red). The fractional number next to each profile indicates the rotational phase. }
    \label{fig:stokes_i_v}
\end{figure*}

We analyzed the ESPaDOnS spectropolarimetric data to study the magnetic field properties of HQ Tau.
We used the Least Square Deconvolution method (LSD) from \cite{Donati97} to compute the mean Stokes I and V photospheric line profiles.
This method increases the S/N by extracting the Zeeman signature in many photospheric lines and averaging them. 
The parameters used for the LSD computation are the mean wavelength, intrinsic line depth, and Land\'e factor, set at 640 nm, 0.2, and 1.2, respectively, as in \cite{Donati10}.
We extracted a list of spectral lines from VALD atomic database for a star with HQ Tau's fundamental parameters, and selected absorption lines located between 450 and 850 nm.
We then removed lines contaminated with emission, blended with strong broad lines, or affected by telluric lines.
The LSD Stokes I and V profiles are shown in Fig. \ref{fig:stokes_i_v}.
The typical rms is $3.5 \times 10^{-4}$ and $1.4 \times 10^{-4}$ for the Stokes I and V profiles, respectively,  computed from more than 9000 photospheric lines. 

     \begin{figure}
        \centering
        \includegraphics[width=0.45\textwidth]{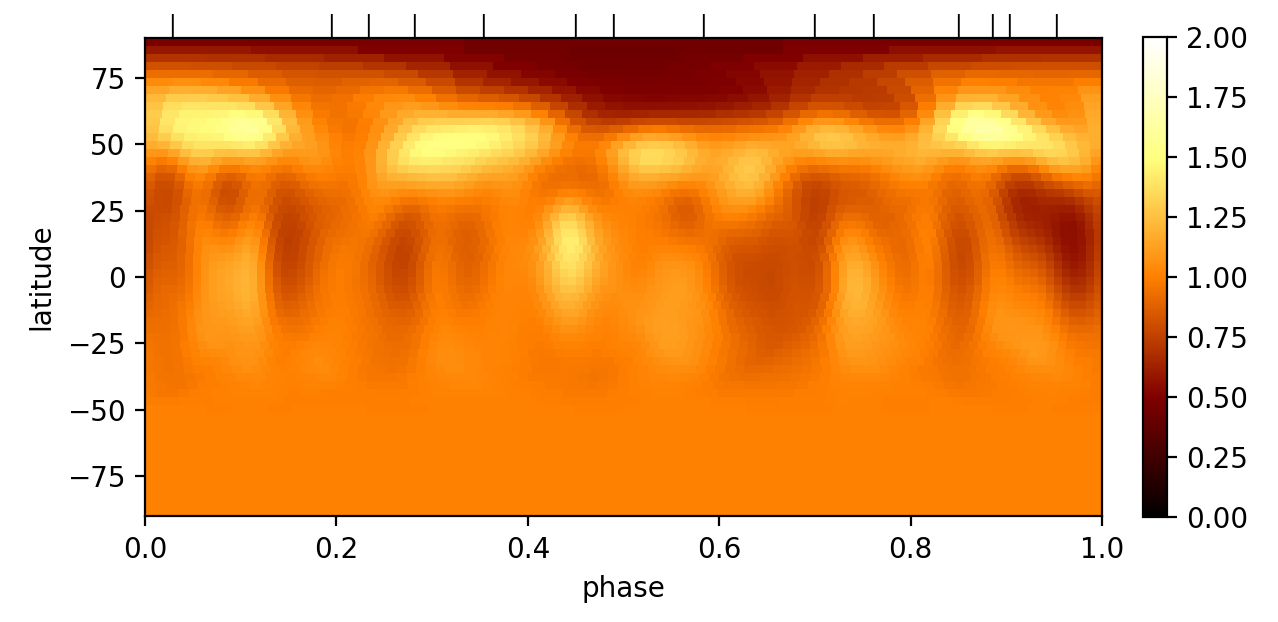}
                \includegraphics[width=.45\textwidth]{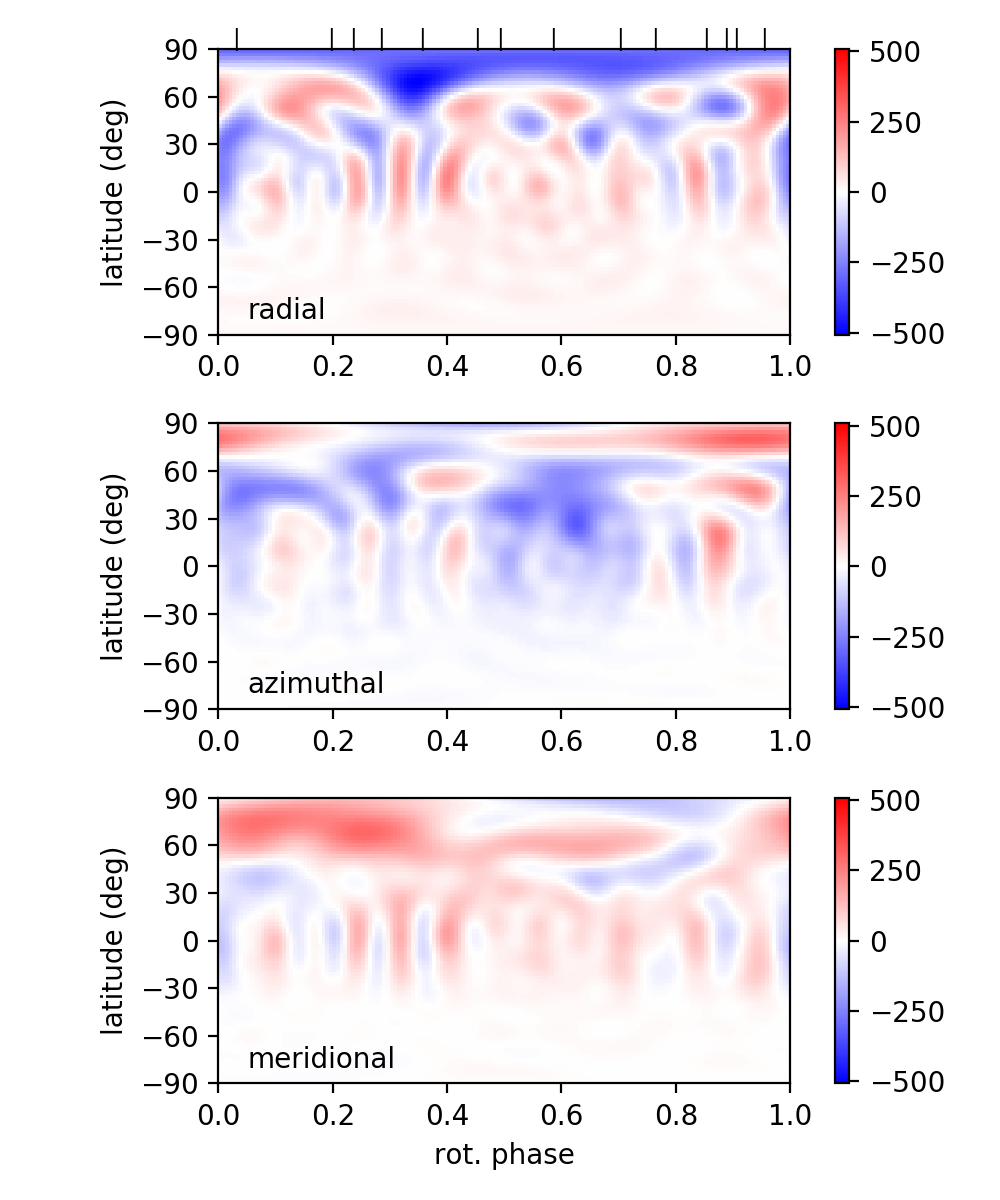}
        \caption{{\it Upper panel:} Surface brightness map of HQ Tau. The color scale goes from dark brown to light yellow, for dark to bright features. {\it Lower panels, from top to bottom:} Radial, azimuthal, and meridional surface magnetic field distribution. Ticks on top of the panels indicate the phase of observations. } 
        \label{fig:bm}
    \end{figure}
        
The shape of the Stokes I profiles displays a clear modulation over the rotational cycle.
A periodogram analysis shows the same peak at a frequency 0.4 d$^{-1}$ as for emission lines, ranging from -25 \kms\ to 60 \kms, with a FAP reaching 10$^{-4}$. This indicates that the modulation is periodic at the stellar rotational period, as expected from a large rotating dark spot at the stellar surface. The profile is nearly symmetric at phase 3.69, confirming the spot is facing the observer around phase 0.65, as defined above from the \vrad\ curve (see Sect. 3.3.1). 
The Stokes V signal is much weaker than the Stokes I profile. Nevertheless, a clear magnetic signature is seen at several phases, with a complex shape. 

The average surface longitudinal magnetic field, $B_l$, was computed from the Stokes I and V profiles, as:

\begin{equation}
    B_l = -2.14 \times 10^{11} \times \frac{\int vV(v)dv}{\lambda g c \int (1-I(v))dv},
    \label{eq:bl}
\end{equation}
    where $B_l$ is given in Gauss, $v$ is velocity relative to line center, $\lambda$ in nm is the mean wavelength, and $g$ the Land\'e factor chosen for the LSD computation \citep{Donati97, Wade00}.
    We computed the error by propagating the uncertainties in the trapezoidal integration over the range $\pm$~60 \kms.

    Figure \ref{fig:param-mod} shows the variation of the longitudinal magnetic field as a function of time and rotational phase. 
    B$_l$ varies from -70 G to +68 G.  We notice that the minimum of B$_l$ is reached at $\phi \sim 0.7$, consistent with the rotational phase at which the dark spot that modulates the radial velocity faces the observer, and reaches a maximum around phase 0.9 when the IPC components are best seen in the emission line profiles. 
    Apart from these extrema, there is little evidence for B$_l$ to be  modulated, possibly the result of a complex magnetic topology at the stellar surface and (or) the strongest field regions being hidden in dark spots. We ran a similar analysis in the CaII IRT 854.2~nm line profile. However, we did not detect any significant Stokes V signature within the line profile and  our data can only provide a 3$\sigma$ upper limit of 450~G to the longitudinal magnetic field component in the line. 
   
    The full Zeeman Doppler Imaging \citep[ZDI,][]{Donati11b, Donati12} analysis was performed based on the LSD Stokes I and V profiles, using the code described in \cite{Folsom18}. The resulting best fit of the Stokes I and V profiles is shown in Fig.~\ref{fig:stokes_i_v}. 
    
    This first step provides a Doppler image shown in Fig.~\ref{fig:bm}. The fitting process starts from a uniform brightness distribution across the stellar disk and iteratively adds darker and brighter features onto the stellar surface to reproduce the shape of Stokes I profile. For each cell on the stellar surface, the local profile is assumed to be a Voigt profile and the brightness of the pixel is adjusted by minimizing $\chi^2$ and maximizing entropy. The solutions for brightness and magnetic field distributions are not unique  based on $\chi^2$. Therefore, an additional constraint is  added using entropy. ZDI uses the algorithm of \cite{Skilling84} for maximizing entropy while minimizing $\chi^2$.
    We adopted 0.66, 2.8 \kms, and 1.9 \kms\ for the limb darkening coefficient, the Gaussian and Lorentzian widths of the Voigt profile, respectively. The latter values were chosen based on theoretical values found in \cite{Gray05} for a K0 spectral type star around 600 nm and adapted to fit the observed profiles. The Doppler image reveals a large cold spot extending over phases 0.4 to 0.8 in longitude, ranging from 90$^{\circ}$ to about 50$^{\circ}$ in latitude, and covering about 20\% of the stellar visible hemisphere. 
    
     ZDI then takes the brightness map as input and fits the Stokes V profiles by adjusting the spherical harmonic coefficients that describe the magnetic field \citep[see][]{Donati06}. It is also possible to fit both intensity and polarisation profiles simultaneously but the resulting maps are essentially the same. Taking advantage of the large \vsini\ of the star, which provides spatial resolution on the stellar surface,  the spherical harmonic expansion was carried out to the 15th order in $l$ \citep{Folsom16}.  Letting the stellar parameters freely vary in the fitting process, we obtain an inclination of 50 $\pm$ 5 $\degr$, a period of 2.453$^{+0.006}_{-0.008}$ d, and a \vsini\ of 51.2 $\pm$ 0.5 \kms. All uncertainties were obtained using the 68.27 \% confidence level on $\chi^2$. All these values are within 3$\sigma$ of those we derived in the previous sections, although the inclination we derive from ZDI is on the lower side of the estimate we obtained in Section 3.2 by combining the rotational velocity and period with the stellar radius. This suggests that the stellar radius may be underestimated by about 20\%. The solution that minimizes $\chi^2$ did not allowed us to constrain the differential rotation d$\Omega$. We thus fixed d$\Omega$ = 0.0 rad d$^{-1}$ to derive the other parameters, as the derived period was consistent within 3$\sigma$ with the photometric one.
            
    The large-scale magnetic reconstruction we recover from the analysis of LSD profiles reveals a mostly poloidal field, with the toroidal component contributing 25\% to the total magnetic energy.
    The main structure of the radial magnetic field extends from phase 0.3 to 0.8, from 90$^{\circ}$ to 60$^{\circ}$ in latitude, and reaches a strength of -562 G locally.
    The dipolar field amounts to only 14\% of the poloidal component, is tilted by 31.5$^{\circ}$ from the rotation axis, and reaches a maximum strength of 63~G at the stellar photosphere. 
    The quadrupole and octupole components contribute respectively 9.7\% and 8.3\% to the total magnetic flux. The complex topology and moderate strength of HQ Tau's magnetic field is reminiscent of those reported by \cite{Hussain09} and \cite{Villebrun19} for IMTTs.

    We caution that the values reported here from the LSD analysis are lower limits to the strength of the  magnetic field that interacts with the disk. Previous similar studies have shown that the ZDI analysis of emission lines arising from the accretion shock located at the foot of funnel flows, most notably HeI 588~nm, yields a much stronger magnetic field strength than that deduced for the large-scale field at the photospheric level from LSD analysis \citep[e.g.,][]{Donati19}. Unfortunately, HQ Tau's spectrum does not show the post-shock HeI line in emission, owing to the modest accretion rate producing a relatively weak shock seen against a bright photosphere.


\section{Discussion}
\label{sec:discussion}

    We selected HQ Tau among many Taurus sources observed by K2 \citep{Rebull20} for a spectropolarimetric follow-up campaign in order to investigate the magnetospheric accretion process in a representative member of the class of relatively massive T Tauri stars, the so-called IMTTs \citep{Calvet04}. With a mass of M$_{\star}$ = 1.9 M$_{\odot}$, a short rotational period of 2.424~d and v$\sin$i = 53.9 \kms, HQ Tau's properties are intermediate between the cool low-mass T Tauri stars and the hotter intermediate-mass Herbig Ae stars. Indeed, PMS models suggest HQ Tau's interior is already partly radiative, with \rrad/\rstar $\sim$ 0.51, as the star is transiting from the Hayashi to the Henyey track in the HR diagram. The mass accretion rate we derive, \macc $\sim$ 1.16 10$^{-9}$ \msunyr, is however relatively low and more typical of low-mass TTS \citep[e.g.,][]{Mendigutia11}. Only few such IMTTs had been monitored for their magnetic properties so far \cite[e.g.,][]{Hussain09} and the goal here was to explore the possible extension of the magnetospheric accretion process that is ubiquitous among low-mass T Tauri stars to the higher mass range. 
    
HQ Tau is undoubtedly a member of the Taurus star forming region \citep{Luhman18}. Yet, the mean radial velocity we measured, <\vrad> = 7.22 $\pm$ 0.27 \kms, is significantly different from the radial velocity distribution of Taurus members, <\vrad >=16.3 $\pm$ 6.43 \kms\ \citep{Bertout06}.  This led us to suspect that HQ Tau might not be a single source, as usually assumed, and the small but regular drift in \vrad\ seen during the ESPaDOnS run may be additional evidence for that (see Fig.~\ref{fig:param-mod}). Historically, \cite{Simon87} reported the system to be a tight binary from lunar occultation, with a separation of 4.9 $\pm$ 0.4 mas, later revised to 9.0 $\pm$ 2 mas by \cite{Chen90}. However, the former authors state that "the binary nature of HQ Tau is not obvious" and indeed this result was not confirmed by following studies \citep{Richichi94, Simon95, Simon96, Mason96}.  

\begin{figure}
        \centering
        \includegraphics[width=0.45\textwidth]{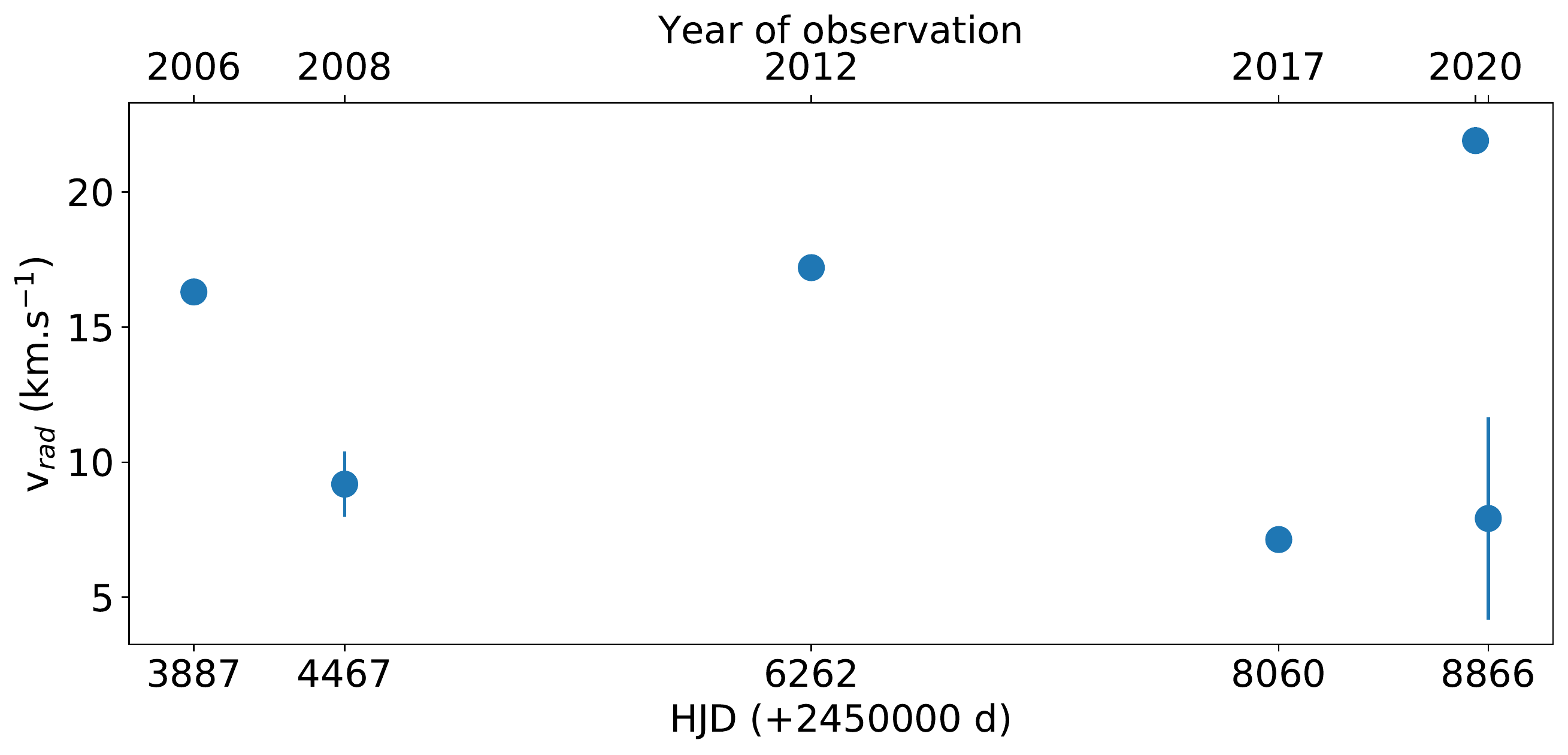}
        \caption{The radial velocity of HQ Tau as a function of date of observation. The error bars are indicated, and can be smaller than the symbol size.}
        \label{fig:all_vrad}
    \end{figure}

We searched for additional radial velocity measurements in the literature. They are summarized in Table~\ref{tab:all_vrad}, together with new measurements, kindly obtained for us recently by L. Hillenbrand and H. Bouy.  The results are illustrated in Fig.~\ref{fig:all_vrad}. HQ Tau exhibits clear velocity variations.  Measurements obtained from 2006 to 2017 show \vrad\ oscillations between about 7 and 22 \kms\ on a timescale of years. However, the two most recent measurements taken 50 days apart in late 2019, show the same amplitude. This suggests that HQ Tau is a short period type 1 spectral binary (SB1), whose orbit determination awaits additional measurements. We note that the companion must be relatively faint compared to the primary in the optical, as we see no evidence for a double-lined system in the Stokes I LSD profiles. Nevertheless, the detection of a secondary period at 5.03~d in the K2 light curve (see Section 3.1) suggests that the companion may account for a low-level contribution to the optical flux of the system. We caution that we have implicitly assumed in the above analysis that the companion's contribution to the emission line flux and to the ZDI reconstruction process could be neglected. Although this assumption may not be fully verified, we have currently no way to estimate the  flux contribution of the companion to the system. Finally, we suggest that the faint mm flux of HQ Tau's disk and its suspected inner cavity \citep{Long19, Akeson19} could conceivably be related to a low-mass companion orbiting within the inner disk and, at least partly, clearing it. 

\begin{table}
    \centering
    \begin{tabular}{lllll}
        \hline
        HJD & \vrad & $\sigma$\vrad & Ref & Instrument \\
        (-2,450,000) & \kms & \kms & - & - \\
        \hline
      3887.77 & 16.3 & 0.02 & 1 & MIKE \\
        4467.83 & 9.2 & 1.2 & 2 & HIRES\\
        6262.77 & 17.2 & 0.2 & 3 & HIRES \\
        8060.00 & 7.22 & 0.15 & 4  & ESPaDOnS\\
        8816.77 & 21.9 & 0.5 & 2 & HIRES \\
        8866.44 & 7.92 & 3.74 & 5 & SOPHIE\\
        \hline
    \end{tabular}
    \caption{\vrad\ values from the literature, and from our and new measurements.}
    \tablebib{(1) \cite{Nguyen12}; (2) L. Hillebrand (priv. comm.); (3) \cite{Pascucci15}; (4) This work; (5) H. Bouy (priv. comm.)}
    \label{tab:all_vrad}    
\end{table}

On a much shorter timescale of days, the radial velocity of the system is modulated by surface spots with a period consistent with the photometric period of 2.424~d derived from the K2 light curve. From the additional ASAS-SN photometry obtained at the time of the spectroscopic campaign, we find that the photometric minimum occurs around $\phi$=0.54, which is close to the rotational phase when the spot which modulates radial velocity variations faces the observer (at $\phi$=0.6 in Fig.\ref{fig:param-mod}). We deduce that a large, cool spot is mainly responsible for both the photometric and radial velocity variability on a timescale of a few days. From the flat-bottom shape of Stokes I LSD profiles at phase 0.69, the relatively low amplitude of variability ($\sim$5\% in both photometric flux and \vrad), and the sinusoidal shape of the K2 light curve, we further infer that the cool spot is presumably situated at a high latitude on the star. Indeed, the Doppler image reconstruction indicates the presence of a cool spot straddling the stellar pole and covering about 20\% of the stellar surface.  

In order to check if a cold spot can produce the observed modulation of radial velocity and photometric flux, we used SOAP2 \citep{Dumusque14} to generate synthetic variability curves. We used the stellar parameters of HQ Tau reported above and ran a set of simulations for a range of spot latitudes, spot sizes, and temperature differences between the spotted area and the stellar photosphere. The strong degeneracy between the spot location and its size does not allow us to strictly constrain these parameters. Nevertheless, we found a solution reproducing the amplitude of modulation of both radial velocity and flux with a spot located at a latitude of 60$\degr$ and whose projected surface covers 30\% of the visible hemisphere, qualitatively consistent with the Doppler map. These spot properties are not uncommon for T Tauri stars \citep[e.g.,][]{Bouvier89}. 

Most of the circumstellar diagnostics also seem to be modulated at the stellar rotation period, consistent with the expectations for magnetospheric accretion. The line profiles exhibit periodically modulated IPC components, whose maximum occurs around phase 0.9, and last for nearly half of the rotational cycle from phase 0.69 to 0.02. This is consistent with magnetospheric accretion funnel flows covering a significant azimuthal extension around the central star, as expected for a tilted large-scale magnetosphere. The shape of the profiles themselves are reminiscent of those computed from magnetospheric accretion models. For instance, the mean \hb\ profile of HQ Tau is quite similar to that computed by  \cite{Muzerolle01} for \macc = 10$^{-9}$ \msunyr\ at an inclination of 60$\degr$, which shows a pronounced IPC component for a funnel flow temperature of 8000~K. The shape of the \ha\ profile is comparable to that computed by \cite{Lima10} for this mass accretion rate. 

    The apparent mass accretion rate itself is modulated at the stellar rotational period. As it scales with the equivalent width of the line profiles, this is mostly the result of the periodic appearance of IPC components, which reduce the line flux. Hence, \macc\ variations are primarily related to the geometric projection of the corotating accretion funnel flow. We do not have evidence for significant intrinsic variations of \macc\ over a timescale of a couple of weeks, although we note that the depth of the IPC components slightly varies over this timescale. The deepest IPC components occur around $\phi = 0.9$, which is significantly later than the epoch at which the cold spot that modulates the radial velocity faces the observer. If the cold spot marks the magnetic pole where the funnel flow is anchored, it might indicate that the upper part of the funnel flow that produces the redshifted absorptions is trailing. This may occur if the magnetospheric truncation radius is located beyond the disk's corotation radius. 
    
    Such a phase delay has been previously reported for T Tauri stars, such as AA Tau \citep{Bouvier03} and V2129 Oph \citep{Alencar12}. Unfortunately, the veiling is too weak in HQ Tau to trace the location of the accretion shock from the modulation of optical excess flux. Indeed, a weak veiling is not unexpected against a relatively bright photosphere (\teff = 4997~K) at such a low \macc. Similarly, we do not see a clear modulation of the longitudinal component of the magnetic field, presumably due to its complex topology, which prevents us from assigning a rotational phase to the magnetic pole. It may therefore be that the accretion shock at the base of the funnel flow is located at a slightly different longitude than the cold dark spot around the stellar pole, thus accounting for the IPC components appearing around $\phi$=0.9 without having to resort to a twisted magnetospheric accretion column.

Together with evidence for magnetospheric funnel flows, some line profiles reveal signatures of outflows. The \ha\ line profile exhibits a high velocity blueshifted absorption component, from about -200 to -150 ~\kms, which is modulated at the stellar rotational period. Interestingly, this component appears to be anti-correlated with the appearance of IPC features in the line profile. This is reminiscent of the "egg beater" model introduced by \cite{Johns95a} for SU Aur, where funnels flow and mass outflows occur on opposite azimuths at the disk inner edge: while accretion funnel flow is favored where the magnetic pole is inclined toward the disk, the situation is reversed at the opposite azimuth and favors interface or inner disk winds. This interpretation would be consistent with the overall variability reported here for the \ha\ profile of HQ Tau. 

Transient mass loss episodes are seen on a timescale of a few days in the NaI D2 line profile. Several deep absorption components are seen over blueshifted velocities ranging from about -40 down to -120~\kms. These absorptions are not rotationally modulated. Instead, they are seen over a few consecutive days and then disappear altogether (see Fig.~\ref{fig:prof_na}). These episodic mass outflows could conceivably be related to magnetic reconnections at the star-disk interface, following the magnetospheric inflation scenario put forward by \cite{Bouvier03} for AA Tau: as the magnetospheric field lines inflate under the shear of differential rotation, they eventually open up and reconnect \citep{Goodson97}. Such an inflationary cycle first induces a phase of reduced mass accretion onto the star during inflation, followed by a transient outflow during reconnection, and the restoration of funnel flow accretion. 

A detailed analysis of the line profile variability may support the magnetospheric inflation scenario. The transient blueshifted absorptions are most conspicuous in the NaI D2 line profile over 2 rotational cycles, from J.D. 8060 to 8065 (see Fig.~\ref{fig:prof_na}), and the central velocity of these components appear to slowly drift toward the line center over this timeframe.  The IPC components seen in Balmer lines are the deepest on J.D. 8066 ($\phi$=4.88, 4.90), meaning right after the transient outflow episode. This would be consistent with the restoration of a magnetospheric accretion funnel after reconnection of  the inflated magnetosphere. In contrast, at the start of the transient outflow episode, on J.D. 8061 ($\phi$=2.85), the IPC profiles are the weakest, consistent with an inflated state for the magnetosphere. As discussed in \cite{Bouvier03} and \cite{Alencar18}, the timescale for the magnetospheric inflation cycle is expected to be several times the rotational period of the star \citep{Zanni13}, which is presumably the reason why it is rarely observed in spectroscopic time series extending over only a couple of weeks.

Although we detect only relatively weak Stokes V signatures in the LSD spectra of HQ Tau, they are sufficient to derive the longitudinal component of the magnetic field and reconstruct surface magnetic maps.  We do not detect rotational modulation of the longitudinal magnetic field, unlike what is usually seen in T Tauri stars \citep[e.g.,][]{Donati19, Donati20}. The rotational modulation of B$_l$ in photospheric lines is often complex, except in the dipolar magnetic field case. It may also be that the weakness of the signatures and the complexity of the field topology at the stellar surface combine to hide the modulation. As HQ Tau has already developed a significant radiative core, we do not expect a surface magnetic field dominated by a dipolar component \citep{Gregory12}. 

The full ZDI analysis suggests a weak dipolar component, inclined by 31.5$\degr$ from the rotational axis, and amounting to about 63~G.  
However, cancellation of opposite polarities on small spatial scales at the stellar surface may affect the total magnetic energy of the star. Also, strong field regions may be hidden in dark stellar spots. Therefore, the total magnetic energy we measure with ZDI on photospheric LSD profiles is to be taken as a lower limit \citep{Lavail19, Sokal20}. Indeed, the magnetic field strength derived from Stokes V signatures in the CaII and (or) HeI emission line profiles of T Tauri stars usually exceeds that deduced from LSD profiles by at least a factor of 5 to 6 because they probe small and localised highly magnetised regions on the surface of the star \citep{Donati19, Donati20}. In HQ Tau, the only emission line available to investigate additional magnetic tracers is CaII IRT. As this line is formed at least partly in the accretion funnel flow, it is a potential probe of the specific field connection between the star and the disk. From the lack of detectable Stokes V signal, we could merely derived a 3$\sigma$ upper limit of $\sim$450~G for the longitudinal field component in the emission line core.  

The two intermediate mass T Tauri stars previously imaged in this way, CR Cha (\prot = 2.3~d) and CV Cha (\prot = 4.4~d) by \cite{Hussain09}, have similar spectral types and masses as HQ Tau but have larger radiative cores (M$_{rad}$/M$_{\star}$ ~$\approx$ 0.65 and 0.92 respectively) and are therefore probably further along in their PMS evolution. In fact, HQ Tau and CR Cha also have very similar rotation periods and mass accretion rates, with \macc = 2~10$^{-9}$ \msunyr\ for  CR Cha \citep{Nisini18}. 
It is interesting to note that there are strong similarities between all three stars in their surface activity, as shown by comparing their brightness maps and their large scale magnetic field distributions. All three stars show a large cool spot near their poles, with CR Cha showing an almost-identical off-centre-polar cap like HQ Tau. The large scale magnetic fields of all three systems are all complex and non-axisymmetric.

Are our results consistent with magnetospheric accretion being at work in HQ Tau? Unfortunately, due to the lack of Zeeman signatures in emission lines, we cannot derive from our data the strength of the dipolar magnetic field in the acretion shock, which would allow us to compute the magnetospheric truncation radius, \rmag, according to \cite{Bessolaz08}'s prescription:
\begin{equation} 
{\frac{r_{mag}}{R_{\star}}}  = 2 m_s^{2/7} B_{\star}^{4/7} \dot{M}_{acc}^{-2/7} M_{\star}^{-1/7} R_{\star}^{5/7}, 
\end{equation}
where m$_s\approx1$, B$_{\star}$ is the equatorial magnetic field strength, 
    \macc\ is the mass accretion rate, \mstar\ the stellar mass, and \rstar\ the stellar radius, respectively in units of 140 G, 10$^{-8}$~\msunyr, 0.8~\msun, and 2~\rsun. Instead, we may assume that the truncation radius is located close to the corotation radius, \rcor = 3.55 $\pm$ 0.35~\rstar. We then derive from Eq.(5) a magnetic field  strength of $\sim$120~G at the equator, which translates to 240~G at the pole for the dipolar component (and to $\sim$3~G at the truncation radius, following \cite{Gregory11}). Although we cannot measure this component directly, we note that the value required at the stellar surface is not inconsistent with the large-scale ZDI magnetic maps derived above nor with the upper limit we set on the CaII line field strength.     

An independent support to the magnetospheric accretion scenario comes from the maximum velocity of IPC components measured in the emission line profiles, $v_{IPC}^{max} \sim$ 330 $\pm$ 19~\kms. We measured this velocity by fitting the red part of H$\beta$'s IPC components by a straight line. The location where the line joins the continuum yields to $v_{IPC}^{max}$. This provides an estimate of the free-fall velocity of the accreted material projected onto the line of sight. 
 Accounting for projection effects that combines the star's inclination ($i\sim 60 \degr$, as the average between the rotational and ZDI estimates above) and the magnetic obliquity ($\beta$=31.5$\degr$), we have $v_{IPC}^{max} \simeq  v_{ff} \cos (i - \beta)$, which yields $v_{ff}$ = 376~\kms. Assuming free fall accretion from \rcor\ to \rstar, the material would hit the stellar surface at a velocity $v_{ff}$ = $(2 G M_\star / R_\star)^{1/2} (1-R_\star/R_{cor})^{1/2}$ = 424~\kms, which is consistent with the estimate obtained from the maximum redshifted velocity of the IPC components, and thus supports a magnetospheric truncation radius being located close to the disk's corotation radius. 

Finally, while we observed HQ Tau during a relatively quiescent phase of variability, we note that its behavior on longer timescales may be more complex. In particular, the deep and long lasting UXOr events reported for the system by \cite{Rodriguez17} must have another origin than the variability described here. The high inclination we derive for the system, $i$ = 75$^{+15}_{-17}$ deg., could favor transient circumstellar extinction events, such as UXOr and (or) AA Tau type.  One of these deep faintening events occurred just before our spectroscopic observations, reaching nearly 1.5 magnitudes and lasting for a month, with significant intra-variability. The system became much redder during the event, consistent with obscuration by circumstellar dust. We may envision two related scenarios to explain this additional component to the system's variability. One is that it might result from a sudden change in the vertical scale-height of the inner disk. As the inner disk edge lies at a distance of only $\sim$3~\rstar\ from the stellar surface, and the system is seen at high inclination, a puffed-up inner disk could conceivably obscure the star for the duration of the instability \citep{Turner10}. Another possibility is a change in the location of the magnetospheric truncation radius relative to the dust sublimation radius. From the expression of \cite{Monnier02}, we derive a sublimation radius \rsub = 5.55 $\pm$ 1.00 \rstar, using Q$_R$ = 1 and T$_s$ = 1500 K. As \rsub\ is larger then \rmag, there is no dust at the magnetospheric truncation radius, which accounts for the lack of dipper-like occultations at the time of our (and K2) observations. However, should the truncation radius increase, following either an increase in the magnetic field strength (by a factor of 3) or a decrease in the mass accretion rate (by a factor of 8), to eventually reach the sublimation radius, a dusty disk warp would result and could induce occultation events in this highly inclined system. We therefore suggest that the episodic UXOr events regularly observed in this system are due to the combination of close-in circumstellar dusty material around the central star and the specific geometry under which the system is seen.


\section{Conclusions}
\label{sec:conclusion}

Following the K2 Taurus campaign, we monitored the young, intermediate mass HQ Tau system with spectropolarimetry in order to investigate the accretion and ejection processes on a timescale of days. The results of this campaign provide clear diagnostics of magnetically mediated accretion occurring in the system. We observed redshifted absorptions in emission line profiles that are periodically modulated at the stellar rotational period, indicative of magnetospheric funnel flows passing through the line of sight. This is expected from a global dipolar topology of the stellar magnetic field at few stellar radii, despite the fact that, at the stellar surface, the Stokes V signatures indicate a more complex topology. While the star hosts a modest dipolar magnetic field component, the reduced mass accretion rate and the star's rapid rotation result in the magnetospheric truncation radius being close to the disk's corotation radius, as observed in most T Tauri systems investigated so far. Concurrent with accretion diagnostics, spectral signatures of outflows are clearly seen in line profiles, some being modulated by stellar rotation, others being transient phenomena, possibly resulting from instabilities at the star-disk magnetospheric boundary. We thus conclude that intermediate-mass pre-main sequence systems may undergo similar accretion and ejection processes as lower mass T Tauri stars, including stable magnetospheric accretion funnel flows, in spite of moderate magnetic field strength. Systems seen at high inclination also experience longer term faintening events, which result from the occultation of the inner system by circumstellar dusty material.

 HQ Tau is a fascinating pre-main sequence system, reported here to possibly be a short period spectroscopic binary whose orbit remains to be determined. It exhibits a mix of variability behaviors, including accretion signatures modulated on a timescale of a few days as seen in lower mass T Tauri stars, as well as longer term UXOr events more typical of Herbig Ae stars. The richness of variability patterns this intermediate-mass young star displays warrants additional multi-wavelength studies of the system and its circumstellar environment on the long term.

\begin{acknowledgements}
      We thank an anonymous referee whose comments improved the content of this paper.
      We thank Lynne Hillenbrand and Herv\'e Bouy for taking additional high resolution spectra of HQ Tau that helped us assessing its binary nature. We acknowledge with thanks the variable star observations from the AAVSO International Database contributed by observers worldwide and used in this research.  This paper includes data collected by the Kepler mission. Funding for the Kepler mission is provided by the NASA Science Mission directorate.
      This work has made use of the VALD database, operated at Uppsala University, the Institute of Astronomy RAS in Moscow, and the University of Vienna. This project has received funding from the European Research Council (ERC) under the European Union's Horizon 2020 research and innovation program (grant agreement No 742095 ; SPIDI : Star-Planets-Inner Disk- Interactions; http://www.spidi-eu.org).
      This project was funded in part by INSU/CNRS Programme National de Physique Stellaire and Observatoire de Grenoble Labex OSUG2020.
      S.H.P. Alencar acknowledges financial support from CNPq, CAPES and Fapemig.
\end{acknowledgements}

\begin{appendix}
\section{Raw emission line profiles}
\begin{figure}[h]
        \centering
        \includegraphics[width=0.45\textwidth]{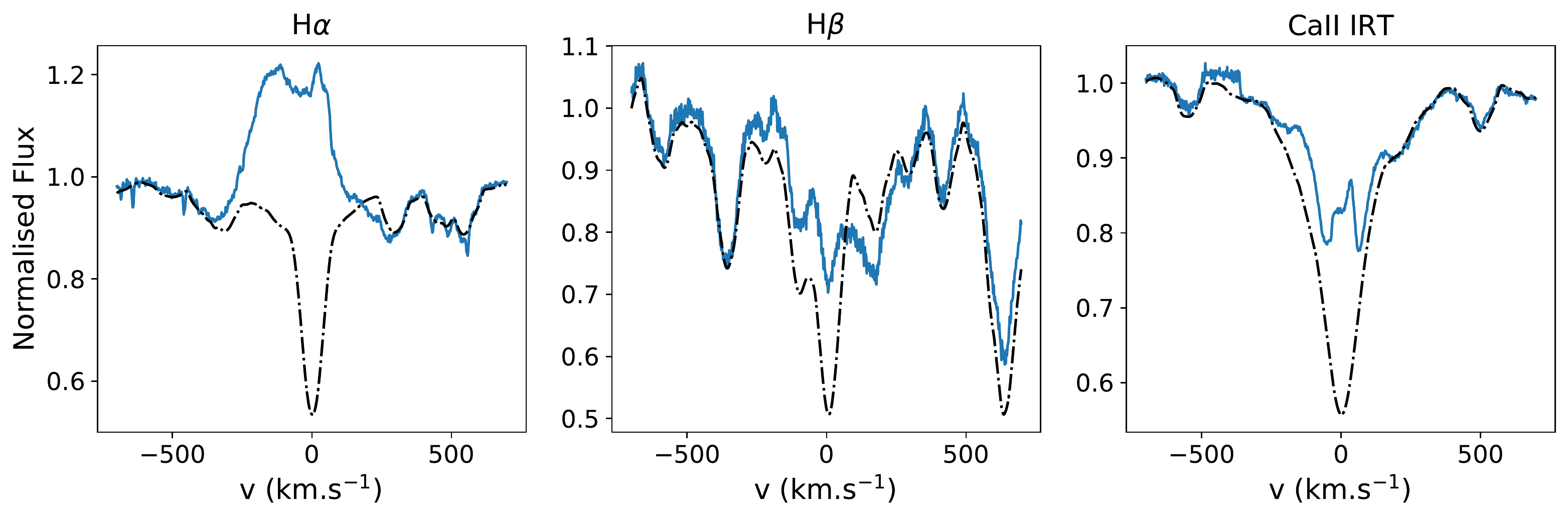}
        \caption{{\it From left to right}: Mean \ha, \hb, and CaII IRT HQ Tau's line profiles (blue) overplotted on those of the photospheric template Melotte~25-151 (dash dotted black). The template has been corrected for the radial velocity shift and rotationally broadened to the \vsini\ of HQ Tau.}
        \label{fig:temp}
    \end{figure}

The average \ha, \hb, and CaII IRT line profiles of HQ Tau are shown in Fig.~\ref{fig:temp}. These figures show the lines before subtracting the photospheric template profiles. 

\end{appendix}

\bibliographystyle{aa}
\bibliography{hqtau}

\end{document}